\documentclass[nofootinbib,prd,12pt,reprint,longbibliography]{revtex4-1}

\usepackage{graphicx}
\usepackage{epstopdf}
\usepackage{amsmath}
\usepackage{amssymb}
\usepackage{amsfonts}
\usepackage{amsthm}
\usepackage{color}
\usepackage{array}
\usepackage{verbatim}
\usepackage{url}
\usepackage{graphicx}
\usepackage{braket}
\usepackage{epsfig}
\usepackage[
      colorlinks=true,
      linkcolor=blue,
      urlcolor=magenta,
      filecolor=blue,
      citecolor=red,
      pdfstartview=FitV,
      pdftitle={},
      pdfauthor={},
      pdfsubject={},
      pdfkeywords={},
      pdfpagemode={},
      bookmarksopen=true
	  ]{hyperref}

\newcommand{\abs}[1]{\left|#1\right|}

\def\coth{{\mathrm{coth}}}
\def\sinh{{\mathrm{sinh}}}
\def\cosh{{\mathrm{cosh}}}

\def\w{{\omega}}

\def\frac#1#2{{#1\over #2}}

\def\b{{\beta}}

\def\k{{\kappa}}

\def\p{\partial}

\def\d{{\mathrm{d}}}

\def\be{\begin{equation}}
\def\ee{\end{equation}}
\def\ba{\begin{eqnarray}}
\def\ea{\end{eqnarray}}

\numberwithin{equation}{section}

\begin{document}

\title{Scalar perturbations of black holes in Jackiw-Teitelboim gravity\\}

\author{Srijit Bhattacharjee$^{a}$\footnote{srijuster@gmail.com}}
\affiliation{${^a}$Indian Institute of Information Technology (IIIT), Allahabad, \\
Deoghat, Jhalwa, Uttar Pradesh, India 211015}

\author{Subhodeep Sarkar$^{a}$\footnote{subhodeep.sarkar1@gmail.com }}
\affiliation{${^a}$Indian Institute of Information Technology (IIIT), Allahabad, \\
	Deoghat, Jhalwa, Uttar Pradesh, India 211015}
	
	\author{Arpan Bhattacharyya$^{a}$\footnote{abhattacharyya@iitgn.ac.in}}
\affiliation{${^a}$Indian Institute of Technology, Gandhinagar, \\ Gujarat, India 382355}

\date{\today}

\begin{abstract}
We study linear scalar perturbations of black holes in two-dimensional (2D) gravity models with a particular emphasis on Jackiw-Teitelboim (JT) gravity. We obtain an exact expression of the quasinormal mode frequencies for single horizon black holes in JT gravity and then verify it numerically using the Horowitz-Hubeny method. For a 2D Reissner-Nordstr\"om like solution, we find that the massless scalar wave equation reduces to the confluent Heun equation using which we calculate the Hawking spectra. Finally, we consider the dimensionally reduced Ba\~nados-Teitelboim-Zanelli (BTZ) black hole and obtain the exterior and interior quasinormal modes. The dynamics of a scalar field near the Cauchy horizon mimics the behavior of the same for the usual BTZ  black hole, indicating a possible violation of the strong cosmic censorship conjecture in the near extreme limit. However, quantum effects seem to rescue strong cosmic censorship. 
\end{abstract}

\maketitle

\section{Introduction}\label{introduction}
The general theory of relativity (GR) is a remarkably successful theory that describes the most fundamental interaction of this Universe, namely, gravitation. The classical theory however breaks down at certain regimes like singularities inside black holes, the singularity at the beginning of the Universe and so on. It is believed that a quantum theory of gravity will resolve these problems. Although there are promising candidates of quantum gravity but all of them suffer from some limitations. The quantum or semiclassical properties of black holes (BH) also pose a number of puzzles. The information loss problem \cite{Hawking:1976ra} related to Hawking radiation \cite{Hawking:1974sw} is one which has baffled physicists for a long time. Finding a resolutions to these problems in D=4 dimensions by merging quantum mechanics with GR has met with modest success. Many of these questions can, however, be addressed with much more control in lower dimensional models of gravity, such as two-dimensional (2D) gravity. Since the 2D Einstein-Hilbert action is just the Gauss-Bonnet topological term, further structure is introduced to invoke the dynamics in these models. One of the most obvious way to do so in a 2D model is by introducing the dilaton field which naturally arises in various compactifications from higher dimensions \cite{Grumiller:2002nm}. Arguably the most prominent of such models is 
the one due to Jackiw and Teitelboim (JT) \cite{Jackiw:1984je, Teitelboim:1983ux}. In recent times, it has played an important role in conjectured duality with the low energy sector of  Sachdev-Ye-Kitaev (SYK) model \cite{Sachdev:1992fk,kitaev:talk,Maldacena:2016hyu} in the context of anti--de Sitter/conformal field theory (AdS/CFT) correspondence \cite{Maldacena:1997re}.  This model and other 2D dilaton models have been thoroughly investigated in different contexts over the last thirty years \cite{Almheiri:2014cka, Mann:1989gh,Mandal:1991tz, NavarroSalas:1999up, Kettner:2004aw} including the much studied string inspired model suggested in \cite{Callan:1992rs}. Recently, in the context of resolution of information puzzle, there has been a rejuvenated interest in studying black holes in JT like models \cite{Hollowood:2020cou,Almheiri:2019qdq}. JT like models have also been studied in the context of the $AdS_2$ geometry obtained in near horizon limit of extremal black holes \cite{Moitra:2018jqs, Nayak:2018qej,Brown:2018bms,Moitra:2019bub}.  

On the other hand isolated black holes are just theoretical artifacts and black holes are always found in a perturbed state just after their formation. A perturbed black hole responds to the perturbations by emitting gravitational waves, the evolution of which in the intermediate stage, is governed by damped oscillatory signals known as quasinormal modes (QNMs). The QNMs are usually studied by introducing a probe scalar field in the background of a black hole and then investigating its evolution. QNMs and scalar perturbations provide a lot of useful information related to black holes like their stability,
gravitational wave spectrum, Hawking radiation, interaction of the black holes with their astrophysical environment and so on (see \cite{Kokkotas:1999bd, Berti:2009kk,Konoplya:2011qq} for comprehensive reviews). QNMs are also important from the  perspective of AdS/CFT correspondence, as QNMs of a AdS$_{D+1}$ black hole are equal to the poles of the retarded Green's function in the D dimensional boundary CFT at strong coupling limit \cite{Horowitz:1999jd}. 

The purpose of the present work is twofold. First, to study the QNMs or scalar perturbations of black holes in JT gravity and examine some of the properties mentioned above. Secondly, to see the classical and quantum backreaction of the probe scalar field on the inner horizon of black holes in JT gravity. A study of the internal structure should also shed light on the status of strong cosmic censorship in 2D black holes which in turn will reveal any possible pathology in these models. 

In this paper we first study scalar perturbations of a single horizon black hole in pure JT gravity and obtain an analytic expression of its QNMs. We then verify this result numerically using the method proposed by Horowitz and Hubeny \cite{Horowitz:1999jd}. In the next section, we study scalar perturbation of a two-dimensional (2D) version of the Reissner-Nordstr\"om (RN) black hole and compute the Hawking radiation spectrum using the Damour-Ruffini method \cite{Damour:1976jd, Sannan:1988eh}. In the following section, we study a black hole obtained by the dimensional reduction of the Ba\~nados-Teitelboim-Zanelli (BTZ) black hole \cite{Banados:1992wn}. We first determine both the interior and exterior quasinormal modes and study its internal structure near the right Cauchy horizon. The internal structure of this black hole shows similar features that were obtained for the BTZ black hole in \cite{Dias:2019ery, Bhattacharjee:2020gbo}. For a similar discussion using a model without the dilaton field, see \cite{Moitra:2020ojo}. We finally study the nature of the singularity at the left Cauchy horizon and calculate the behavior of the quantum stress energy tensor near the right Cauchy horizon. We then conclude with a discussion of the results and a possible outlook.
\section{Jackiw-Teitelboim Gravity and Quasinormal Modes}
\subsection{The model}
We begin by introducing the most general (1+1)-dimensional action that depends on the metric $\bar{g}_{\mu \nu}$ and a scalar dilaton $\bar{\phi}(r)$, and is compatible with diffeomorphism invariance and contains no more than double derivative of the field \cite{Kettner:2004aw},
\begin{equation}
    S=\dfrac{1}{2G} \int \d x^2 \sqrt{-\bar g}\left( D(\bar{\phi})R(\bar{g})+\dfrac{1}{2}(\nabla \bar{\phi})^2 + \dfrac{V_{\bar{\phi}}(\bar{\phi})}{L^2}\right),
\end{equation}
where $D(\bar{\phi})$ and $V_{\bar{\phi}}(\bar{\phi})$ are model-dependent functions of the dilaton, $G$ is a dimensionless gravitational coupling and $L$ is another fundamental parameter (a length scale) of the two-dimensional theory which will be interpreted as the AdS$_2$ radius.

Demanding that both $D(\bar{\phi})$ and its derivatives are nonvanishing, we can reparametrize the fields in the following way so that the kinetic term gets eliminated from the action \cite{LouisMartinez:1993eh, LouisMartinez:1993cc, LouisMartinez:1995rq}:
\begin{subequations}
\begin{align}
    & \Omega^2(\bar{\phi}) \equiv \exp\left( \dfrac{1}{2}\int \dfrac{\d \bar{\phi}}{(\d D/\d \bar{\phi})}\right), ~  g_{\mu \nu} \equiv \Omega^2(\bar{\phi})\bar{g}_{\mu \nu}, \nonumber \\ &  \phi \equiv D(\bar{\phi}), ~ V(\phi) \equiv V_{\bar{\phi}}/\Omega^2. 
\end{align}
\end{subequations}
Then, the reparametrized action for a generic two-dimensional model is given by 
\begin{equation}
\label{gen2daction}
	S=\dfrac{1}{2G}\int \d^2 x \sqrt{-g}\left( \phi R + \dfrac{V(\phi)}{L^2}\right).
\end{equation}
If we now set 
\begin{equation}
	V(\phi)=2\phi,
\end{equation}
we get the action for Jackiw-Teitelboim (JT)  gravity \cite{Teitelboim:1983ux,Jackiw:1984je} from the above action, namely,
\begin{equation}
\label{JTaction}
	S=\dfrac{1}{2G}\int \d^2 x \sqrt{-g}\phi\left( R + \dfrac{2}{L^2}\right).
\end{equation}
Starting from the Einstein-Hilbert action in $(2+1)$ dimensions with a negative cosmological constant \cite{Banados:1992wn}, we can also obtain the above action by performing a dimensional reduction \cite{Achucarro:1993fd}.

Now, with a gauge choice $\phi=r/L$, we can obtain a Schwarzschild-like solution from the action \eqref{JTaction} \cite{LouisMartinez:1993cc, Gegenberg:1994pv, Achucarro:1993fd}, viz.,
\begin{equation}\label{schbh}
	\d s^2=-f(r)\d t^2+\dfrac{1}{f(r)}\d r^2, ~ \phi=\dfrac{r}{L},
\end{equation}
where 
\begin{equation}
    f(r)=r^2/L^2-2GLM, \label{lapsefuncJT}
\end{equation}
$M$ being the mass of the black hole.
This black hole solution clearly has a horizon at 
\begin{equation}
r_h=L\sqrt{(2GLM)}.
\end{equation}
We can also define the tortoise coordinate $r_*$ using the relation $\d r_*=\d r/f(r)$, and introduce the coordinate $v=t+r_*$ and write \eqref{schbh} in Eddington-Finkelstein coordinates as
\begin{equation}
    \d s^2 = -f(r) \d v^2 + 2 \d v \d r. \label{edfinschbb}
\end{equation}

In the next subsection we will study scalar perturbations and the resultant wave equation in the $2D$ spacetime that we have just described. 
\subsection{The scalar wave equation}
Here we shall take the perturbing field to be a massless scalar field $\Phi(x,t)$. But before studying such perturbations,  we have to ensure that its equation of motion captures the dynamics of a matter perturbation in the two-dimensional black hole background described by \eqref{schbh}. To this end, we generalize the usual Klein-Gordon equation and include an arbitrary coupling, $h(\phi)$, between the dilaton and the matter perturbation and therefore, use the following generalized Klein-Gordon equation \cite{Kettner:2004aw}:
\begin{equation}
\label{genKG}
	\dfrac{1}{\sqrt{-g}h(\phi)}\partial_\mu\left(\sqrt{-g}h(\phi)g^{\mu \nu}\partial_\nu \psi\right)=0.
\end{equation}
The choice of the coupling  $h(\phi)=h(r)$ is inspired by the fact that the metric \eqref{schbh} can be obtained from the dimensional reduction of a BTZ black hole (as we have already noted). It must have the part of the higher dimensional metric-determinant that is lost through dimensional reduction. In our case, this choice simply translates to $h(\phi) =\phi=r/L$ \cite{Kunstatter:1997my, Kettner:2004aw}. It is crucial to take this kind of coupling into account since this coupling determines the form of the potential (as shown below) which plays the key role in the scattering problem. If we do not consider such a coupling then the spacetime will not be able to support massless modes \cite{Cordero:2012je}.

We now consider the following ansatz: 
\begin{equation}
	\Phi(r,t)=\dfrac{R(r)}{\sqrt{h(r)}}e^{-i \omega t},
\end{equation}
and write down wave equation \eqref{genKG} using \eqref{schbh} in a Schrd\"odinger equation-like form
\begin{equation}
    \label{schodingereqJT}
	\partial_{r_{*}}^2R(r)+ \left[ \omega^2-V(r)\right]R(r)=0,
\end{equation}
where we have used the tortoise coordinate $r_*$ and the potential is given by
\begin{equation}
    \label{potentialJT1}
	V(r)=\dfrac{1}{2}\dfrac{f}{h}\left[ f h''+f'h'-\dfrac{1}{2}\dfrac{f}{h}(h')^2\right].
\end{equation}
This is analogous to the Regge-Wheeler equation. Here prime denotes a differentiation with respect to $r$. Using \eqref{lapsefuncJT} and setting $L=1$, such that $h(r)=r$ in the above equation, we get
\begin{align}
        \label{potentialJT2}
	V(r) &= - GM - \dfrac{G^2 M^2}{r^2}+\dfrac{3r^2}{4}, \nonumber \\
	     & =-\frac{r_h^2}{2}-\dfrac{r_h^4}{4r^2}+\dfrac{3r^2}{4},
\end{align}
where $r_h=\sqrt{2GM}$ for L=1 and it is implicit that $r$ is a function of $r_*$. We also note that setting $L=1$ is equivalent to the simultaneous rescalings, $r \to \tilde{r}=r/L$ and $\w \to \tilde{w}=\w L$. Through this rescaling, we measure the frequency and other quantities in terms of the AdS$_2$ radius \cite{Cardoso:2001hn}.

We note that we may also use an ansatz of the form 
\begin{equation}
	\Phi(r,t)=\dfrac{\bar{R}(r)}{\sqrt{h(r)}}e^{-i \omega v}, ~ h(r)=r,
\end{equation}
and write \eqref{genKG}, using \eqref{edfinschbb}, as
\begin{equation}
\label{HH_KG_eq1}
f(r) \partial_r^2 \bar{R}(r)+ [f'(r)-2 i \w] \partial_r \bar{R}(r)-\bar{V}\bar{R}(r)=0,
\end{equation}
where 
\begin{equation}
    \bar{V}=\left[\dfrac{f'}{2r}-\dfrac{f}{4r^2} \right].
\end{equation}
We note $R(r)=e^{-i\w r_*}\bar{R}(r)$. This form of the Klein-Gordon equation will be useful for studying the quasinormal mode frequencies (QNMs) numerically later whereas \eqref{schodingereqJT} will be used to find out exact expressions of QNMs in the next subsection.
\subsection{Quasinormal modes for scalar perturbations}
Quasinormal modes are those mode solutions which are purely ingoing near the horizon of the black hole and vanish at infinity in AdS spacetimes \cite{Kokkotas:1999bd,Berti:2009kk,Cardoso:2001hn,Konoplya:2011qq}. This choice of boundary condition is motivated by the fact that the potential diverges at infinity. In this section we first compute the exact quasinormal mode frequencies and then verify the result numerically.
\subsubsection{Exact calculation}
Before proceeding, we note that we can use the definition of the tortoise coordinate to obtain an implicit form of $r(r_*)$, viz., $r=-r_h\coth(r_h r_*)$ for $L=1$ (therefore $r_* \to -\infty$ corresponds to $r=r_h$ and $r_*=0$ corresponds to $r \to \infty$). Using this, we can rewrite \eqref{potentialJT2} as \footnote{We note that this form of the potential is quite general and arises in a variety of contexts, for example, in pure de Sitter spacetimes \cite{Du:2004jt}. For a discussion on using the analytical continuation approach to compute QNMs, see \cite{Fabris:2020kog}.}
\begin{equation}
	V(r)=\dfrac{3r_h^2}{4\sinh^2(r_hr_*)}+\dfrac{r_h^2}{4\cosh^2(r_hr_*)}.
\end{equation}

So, our wave equation \eqref{schodingereqJT} reduces to
\begin{equation}
	\dfrac{\partial^2 R(r)}{\partial r^2_*}+ \left(\omega^2 - \dfrac{3r_h^2}{4\sinh^2(r_hr_*)}-\dfrac{r_h^2}{4\cosh^2(r_hr_*)}\right)R(r)=0.
\end{equation}

We can now do yet another change of variable, such that it maps the horizon to $x=0$ and the infinity to $x=1$, namely,
\begin{equation}
	x=\dfrac{1}{\cosh^2(r_h r_*)},
\end{equation} 
where $x \in [0,1]$, and rewrite the above equation as
\begin{align}\label{qnmhyp}
&	4x(1-x)\dfrac{d^2R}{dx^2}+(4-6x)\dfrac{dR}{dx} \nonumber \\ & -\dfrac{r_h^2}{4x(1-x)}\left(\dfrac{4 \omega^2 (1-x)}{r_h^2} -3x -x(1-x)\right)=0.
\end{align}
 
We now define 
\begin{equation}
    R(x)=(x-1)^{3/4}x^{-i\w/2r_h}F(x),
\end{equation}
then \eqref{qnmhyp} assumes the standard hypergeometric form \cite{Nikiforov1988,Cardoso:2001hn},
\begin{align}
   & x(1-x)\dfrac{\d^2F(x)}{\d x^2} +\left(c-(1+a+b)x\right)\dfrac{\d F(x)}{\d z} \nonumber \\ & -ab F(x)=0,
        \label{JThypgeo}
\end{align}
with 
\begin{equation}
    a=1-i\dfrac{\w}{2r_h}, ~ b=1-i\dfrac{\w}{2 r_h}~ \mathrm{and}~ c=1-i\dfrac{\w}{r_h}.
\end{equation}
The solutions to \eqref{JThypgeo} are given by the standard hypergeometric function of the second kind (denoted by $_2F_1$). Moreover, the hypergeometric equation has three singular points at $x=0$, $x=1$ and $x=\infty$ and two independent solutions in the neighbourhood of each singular point. Since we are trying to calculate the quasinormal modes, we will impose the boundary condition on the solutions in the range $[0,1]$ such that they are purely ingoing near $x=0$ and vanish at $x=1$. The solution which is purely ingoing at the $x=0$ is $F(x)={_2F_1(a,b;c,x)}$ \cite{Birmingham:2001hc}. We now make use of the following two relations \cite{abramowitz+stegun}
\begin{subequations}
\begin{align}
     &_2F_1(a,b;c,x)=(1-x)^{c-a-b}{_2F_1(c-a,c-b;c;x)}, \\ &_2F_1(c-a,b-a;c;1)=\dfrac{\Gamma(c)\Gamma(a+b-c)}{\Gamma(a)\Gamma(b)}.
\end{align}
\end{subequations}
Applying the boundary condition $F(x)=0$ at $x=1$ is therefore equivalent to imposing the restriction \cite{Cardoso:2001hn},
\begin{equation}
    a=-n ~\mathrm{or}~b=-n,
\end{equation}for $n=0,1,2,\cdots$. This gives us the quasinormal frequencies,
\begin{equation}
\label{JT_QNM}
	\omega=-2i r_h(n+1).
\end{equation}
We note that the QNMs are purely imaginary. We verify this result numerically in the next subsection.
\subsubsection{Numerical calculation}
We now use the method due to Horowitz and Hubeny \cite{Horowitz:1999jd} to numerically compute the quasinormal mode frequencies. We start with \eqref{HH_KG_eq1} and switch to a new variable 
\begin{equation}
    y=1/r,~ h=1/r_h,
\end{equation}
and rewrite \eqref{HH_KG_eq1} as
\begin{equation}
    \label{HH_KG_eq2}
    s(y)\dfrac{\d^2 \bar{R}(y)}{\d y^2}+\dfrac{t(y)}{y-h}\dfrac{\d \bar{R}(y)}{\d y}+\dfrac{u(y)}{(y-h)^2}\bar{R}(y)=0,
\end{equation}
where the coefficient functions are
\begin{subequations}
\begin{align}
    &s(y)=\dfrac{-y^2+y^4/h^2}{y-h}=\dfrac{y^2}{h}+\dfrac{y^3}{h^2}, \\
    &t(y)= \dfrac{2y^3}{h^2}-2 i\w x^2, \\
    &u(y)=(y-h) \bar{V}(y), \\
    &\bar{V}(y)=\left[\dfrac{3}{4}-\dfrac{y^2}{4h^2} \right].
\end{align}
\end{subequations}
Now, \eqref{HH_KG_eq2} has two regular singular points at $y=0$ and $y=h$. So we can use the power series method and look for a solution of the form,
\begin{equation}
    \bar{R}(y)= \sum_{n=0}^{\infty}a_n(y-h)^{n+\alpha}.
\end{equation}
Imposing the boundary condition that we have only purely ingoing modes at the horizon amounts to setting $\alpha=0$ \cite{Horowitz:1999jd, Cardoso:2001hn}. So we look for solutions of the form,
\begin{equation}
\label{hh_sol}
        \bar{R}(y)= \sum_{n=0}^{\infty}a_n(y-h)^n.
\end{equation}
We finally impose the boundary condition that the modes must vanish at infinity ($z=0$) and this amounts to
\begin{equation}
\label{HH_expression_final}
     \sum_{n=0}^{\infty}a_n(y-h)^n=0.
\end{equation}
We can now solve this polynomial equation numerically to determine the roots $\omega$. The coefficients $a_n$ themselves are determined through the recursion relations obtained by substituting \eqref{hh_sol} in \eqref{HH_KG_eq2} (please see \cite{Horowitz:1999jd, Cardoso:2001hn, Cardoso:2001bb, Cardoso:2003cj} for details). Since we cannot determine the full sum in the expression \eqref{HH_expression_final}, we evaluate a partial sum from $0$ to (say) $N$ and find the root $\w$. We then include the next $N+1$ term and determine the roots. If the method is reliable then the roots converge. We have used a modified version of the code made publicly available by Cardoso and his collaborators \cite{Cardoso:2001bb, Cardoso:2003cj}, and we have calculated the roots up to three decimal digit precision and report them here in Table \ref{table:JT_QNM}. We see that the numerical results completely agree with \eqref{JT_QNM}. 
\begin{table}[htpb]
\begin{ruledtabular}
\begin{tabular}{lcr}
$r_h$              & Numerical                &    Theoretical          \\
		          & $\w_N=\w_r+i\w_i$         &  $\w_{th}=\w_r+i\w_i$   \\ 
\hline
		   $1$    &  $-2.000i$                &   $-2i$                  \\ 
		   $5$    &  $-10.000i$               &   $-10i$                 \\ 
		   $10$   &  $-20.000i$               &   $-20i$                 \\ 
	       $50$   &  $-100.000i$              &   $-100i$                \\ 
	       $100$  &  $-200.000i$              &   $-200i$                \\ 
	       $500$  &  $-1000.000i$             &   $-1000i$               \\ 
	   	   $1000$ &  $-2000.000i$             &   $-2000i$               \\ 
\end{tabular}
\caption{\label{table:JT_QNM}Numerical ($\w_N$) and theoretical values ($\w_{th}$) of the lowest quasinormal mode frequencies ($n=0$) for some select black hole sizes ($r_h$). We see that the frequencies are purely imaginary and negative in both cases and match up to three decimal places.}
\end{ruledtabular}
\end{table}

In general, for black holes in 2D Einstein-dilaton systems the quasinormal frequencies usually contain both real and imaginary parts and the real part could possibly be related to the quantized area spectrum of the black hole entropy via the {\it Hod conjecture} \cite{Hod:1998vk,Kettner:2004aw}. However, in this Jackiw-Teitelboim model, we do not get any real part of the QNMs and hence Hod's conjecture is not applicable here. Moreover, the QNMs that we have obtained are related to the relaxation or thermalization time of the dual CFT living at the one-dimensional boundary, and they should coincide with the poles of the two-point retarded Green’s function. In the next section we consider a multihorizon black hole and attempt to study its scalar modes.

\section{The 2D Reissner-Nordstr\"om black hole}
Consider the following two-dimensional action obtained from the dimensional reduction of the 3+1-dimensional Einstein-Maxwell action followed by a suitable reparametrization \cite{NavarroSalas:1999up,LouisMartinez:1995rq,Brown:2018bms}:
\begin{equation}
    S=\dfrac{1}{2} \int \d^2 x \sqrt{-g}\left(\phi R + V(\phi)\right),
\end{equation}
where 
\begin{equation}
    V(\phi)=\dfrac{1}{L^2}\left( \dfrac{1}{\sqrt{2 \phi}}-\dfrac{Q^2}{(2 \phi)^{3/2}}\right).
\end{equation}
Here $L$ is the AdS radius, and $\phi$ is a scalar dilaton field. The general solution with a nonconstant dilaton is 
\begin{equation}
    \d s^2=-f(x)\d t^2+ \dfrac{1}{f(x)}\d x^2,
\end{equation}
where
\begin{equation}
    f(x)=\sqrt{\dfrac{2 x}{L}}-2 L M + Q^2 \sqrt{\dfrac{L}{2x}}.
\end{equation}
This corresponds to a reparametrization of the standard Reissner-Nordstr\"om (RN) solution with the gauge choice $\phi(x)=x/L$ \cite{NavarroSalas:1999up}. Here $M$ denotes the mass of the black hole. This black hole has two horizons located at 
\begin{equation}
    \sqrt{x}_\pm=\frac{\sqrt{L}(LM \pm \sqrt{L^2 M^2-Q^2})}{\sqrt{2}}.
\end{equation}
We note that we can rewrite $f(x)$ in terms of the inner and outer horizons as 
\begin{equation}
    f(x)=\sqrt{\frac{L}{2x}}\left(\sqrt{x}-\sqrt{x}_+\right)\left(\sqrt{x}-\sqrt{x}_-\right),
\end{equation}
and the surface gravity at the outer horizon $\k_+$ is given by
\begin{equation}
\label{RN_k+}
    \k_+=\sqrt{L}\left(\dfrac{\sqrt{x}_+-\sqrt{x}_-}{4\sqrt{2}x_+}\right)
\end{equation}
\subsection{Scalar field perturbations}
We now derive the equation of motion of a massless scalar field, $\Phi(x,t)$, in the black hole background. In order to take into account the coupling of the dilaton with the matter perturbation, we use the generalized form of the Klein-Gordon equation 
\begin{equation}
\label{kg_rn1}
    \dfrac{1}{\sqrt{-g}h(\phi)} \p _\mu \left(g^{\mu \nu} h(\phi) \p_\nu \Phi(x,t)\right)=0.
\end{equation}
We consider a linear coupling, such that, $h(\phi)=\phi(x)=x/L$. Using the ansatz 
\begin{equation}
    \Phi(x,t)=e^{-i \w t}R(x)
\end{equation}
we can write the Klein-Gordon equation as
\begin{equation}
\label{KGRN}
    \dfrac{1}{x} \p_x \left(x f(x) \p_x R(x)\right)+\dfrac{\w^2}{f(x)}R(x)=0.
\end{equation}
We now introduce a new variable $z$ which maps the outer horizon to $z=0$ and the inner horizon to z=1,
\begin{equation}
\label{rn_z}
    z=\dfrac{\sqrt{x}-\sqrt{x}_+}{\sqrt{x}_--\sqrt{x}_+}.
\end{equation}
This transformation enable us to write the Klein-Gordon equation as
\begin{align}
    \label{kg_z}
    & \dfrac{\d^2 R}{\d z^2}+ \left( \dfrac{1}{z}+\dfrac{1}{z-1}\right)\dfrac{\d R}{\d z} \nonumber \\ & +\left( -\tilde{\alpha}_0^2+ \dfrac{\tilde{\alpha}_1}{z} + \dfrac{\tilde{\alpha}_2}{z-1} -\dfrac{\tilde{\alpha}_3^2}{z^2}-\dfrac{\tilde{\alpha}_4^2}{(z-1)^2} \right)=0,
\end{align}
where
\begin{subequations}
\label{tilde_params}
\begin{align}
    \tilde{\alpha}_0^2 &=-\dfrac{8(\sqrt{x}_- -\sqrt{x}_+)^2 \w^2}{L}, \\
    \tilde{\alpha}_1   &=\dfrac{16(2\sqrt{x}_- x_+^{3/2}-x_+^2)\w^2}{L(\sqrt{x}_- -\sqrt{x}_+)^2}, \\
    \tilde{\alpha}_2   &=\dfrac{16(x_-^2-2\sqrt{x}_+ x_-^{3/2})\w^2}{L(\sqrt{x}_- -\sqrt{x}_+)^2}, \\
    \tilde{\alpha}_3^2 &=-\dfrac{8 x_+^2 \w^2}{L (\sqrt{x}_- -\sqrt{x}_+)^2}, \\
    \tilde{\alpha}_4^2 &=-\dfrac{8 x_-^2 \w^2}{L (\sqrt{x}_+ -\sqrt{x}_+)^2}. 
\end{align}
\end{subequations}
We can now choose the parameters $\tilde{\alpha}_0$, $\tilde{\alpha}_3$ and $\tilde{\alpha}_4$ in the following way:
\begin{align}
\label{a3t}
   & \tilde{\alpha}_0=i\dfrac{2\sqrt{2}(\sqrt{x}_+ -\sqrt{x}_-) \w}{\sqrt{L}}, \nonumber \\ &\tilde{\alpha}_3=i\dfrac{2\sqrt{2}x_+ \w}{(\sqrt{x}_+ -\sqrt{x}_-)\sqrt{L}}, \nonumber \\ &\tilde{\alpha}_3=i\dfrac{2\sqrt{2}x_- \w}{(\sqrt{x}_+ -\sqrt{x}_-)\sqrt{L}},
\end{align}
and implement a transformation of following type
\begin{equation}
\label{RNhansatz}
    R(z)=e^{\tilde{\alpha}_0}z^{\tilde{\alpha}_3}(z-1)^{\tilde{\alpha}_4}H(z),
\end{equation}
which changes \eqref{kg_z} to the confluent Heun equation \cite{Fiziev_2009}:
\begin{align}
    \label{Hueneq}
  &  \dfrac{\d^2 H(z)}{\d z^2}+ \left( \alpha +\dfrac{1+\beta}{z}+\dfrac{1+\gamma}{z-1} \right)\dfrac{\d H(z)}{\d z} \nonumber \\ & +\left( \dfrac{\mu}{z} +\dfrac{\nu}{z-1}\right)H(z)=0,
\end{align}
where the standard Heun parameters $\alpha$, $\beta$, $\gamma$ are related to $\tilde{\alpha}_0$, $\tilde{\alpha}_3$ and $\tilde{\alpha}_4$ as 
\begin{equation}
\label{alphabetagamma}
    \alpha = 2\tilde{\alpha}_0, ~ \beta = 2\tilde{\alpha}_3 ~ \mathrm{and} ~\gamma = 2\tilde{\alpha}_4,
\end{equation}
and the parameters $\mu$ and $\nu$ are related to the standard Heun parameters $\eta$ and $\delta$ through the relations \cite{Fiziev_2009},
\begin{align}
    \mu & = \tilde{\alpha}_0+\tilde{\alpha}_1- \tilde{\alpha}_3-\tilde{\alpha}_4+2\tilde{\alpha}_0 \tilde{\alpha}_3 -2 \tilde{\alpha}_3 \tilde{\alpha}_4, \nonumber  \\
        & =\frac{1}{2}(\alpha - \beta -\gamma + \alpha \beta - \beta \gamma) -\eta, \\
    \nu & =\tilde{\alpha}_0+\tilde{\alpha}_2+\tilde{\alpha}_3+\tilde{\alpha}_4+ 2 \tilde{\alpha}_0 \tilde{\alpha}_4+ 2 \tilde{\alpha}_3 \tilde{\alpha}_4, \nonumber \\
        & = \frac{1}{2}(\alpha +\beta +\gamma +\alpha \gamma + \beta \gamma) + \delta +\eta ,
\end{align}
Hence we can deduce that 
\begin{equation}
\label{etadelta}
    \eta = -\tilde{\alpha}_1 ~ \mathrm{and} ~ \delta = \tilde{\alpha}_1+ \tilde{\alpha}_2.
\end{equation}

The confluent Heun equation has a unique particular solution which is regular around the regular singular point $z=0$ (the outer horizon). This solution is known as the standard confluent Heun function $H_C(\alpha, \beta, \gamma, \delta, \eta,z)$ \cite{Fiziev_2009, Bhattacharyya:2020zkc}. 

\subsection{Hawking radiation}
Using the standard confluent Heun function, $H_C(\alpha, \beta, \gamma, \delta, \eta,z)$, the general solution to the radial equation \eqref{KGRN} in the region outside the black hole ($0<z<\infty$) can be written as \cite{Bhattacharyya:2020zkc}
\begin{align}
    R(z)=& e^{\frac{1}{2}\alpha z}z^{\frac{1}{2}\beta}(z-1)^{\frac{1}{2}\gamma} \left(c_1 H_C(\alpha, \beta, \gamma, \delta, \eta,z) \nonumber \right. \\ & \left.  +c_2 z^{-\b}H_C(\alpha, -\beta, \gamma, \delta, \eta,z) \right),
\end{align}
where $c_1$ and $c_2$ are constants, and we have used \eqref{tilde_params}, \eqref{RNhansatz} and \eqref{alphabetagamma}. Moreover, $H_C(\alpha, \beta, \gamma, \delta, \eta,z)$ can be written as a convergent power series about $z=0$, viz.,
\begin{equation}
H_C(\alpha, \beta, \gamma, \delta, \eta,z)=\sum_{n=0}^{\infty} a_n z^n, ~ \abs{x}<1.
\end{equation}
The coefficients can be determined using the three-term recurrence relation given in \cite{Bhattacharyya:2020zkc} and for the present purpose, we note that $a_0=1$. We can then immediately write down the behavior of the radial function $R(z)$ near the outer horizon ($x \to x_+$) \cite{Vieira:2014iea} up to the leading order as
\begin{equation}
    R(x) \sim c_1 (\sqrt{x}-\sqrt{x}_+)^{\b/2}+c_2(\sqrt{x}-\sqrt {x}_+)^{-\b/2}
\end{equation}
where we have used \eqref{rn_z} and have absorbed all the constants into $c_1$ and $c_2$. Therefore, the solution to the Klein-Gordon equation \eqref{kg_rn1} can be written as
\begin{align}
    \Phi(x,t) \sim & e^{-i \w t}(\sqrt{x}-\sqrt{x}_+)^{i \w/2 \k_+} \nonumber \\ & + e^{-i \w t}(\sqrt{x}-\sqrt{x}_+)^{-i \w /2 \k_+},
\end{align}
where we have used $\beta=i{\omega}/{\k_+}$ [cf. \eqref{RN_k+},\eqref{a3t}, \eqref{alphabetagamma}]. Now, we can approximate $f(x)$ near $x_+$ as $f(x)\approx2\kappa_+(\sqrt{x}-\sqrt{x}_+)$, and we can define the usual tortoise coordinate $\d x / \d x_*$ to write
\begin{equation}
    (\sqrt{x}-\sqrt{x}_+) \sim e^{2 \k_+x_*},
\end{equation}
near the outer horizon. Then introducing the usual null coordinate $v=t+x_*$ we can identify the ingoing and outgoing modes as
\begin{subequations}
\begin{align}
 \Phi_{\mathrm{in}} &= e^{-i \w t}(\sqrt{x}-\sqrt{x}_+)^{-i \w /2 \k_+} \nonumber \\ &= e^{-i\w v}, \\
 \Phi_{\mathrm{out}}(x>x_+) &=e^{-i \w t}(\sqrt{x}-\sqrt{x}_+)^{\b/2} \nonumber \\ &= e^{-i\w v}(\sqrt{x}-\sqrt{x}_+)^{i\w/\k_+}. \label{nonanalyticmode}
\end{align}
\end{subequations}
Next, we use the Damour-Ruffini \cite{Damour:1976jd, Sannan:1988eh, Vieira:2014iea, Sakalli:2016fif} method to obtain the Hawking radiation spectra. We see that the mode \eqref{nonanalyticmode} is not analytical in the outer horizon at $x=x_+$, We can therefore perform an analytical continuation by a rotating through $-\pi$ in the lower half complex $x$ plane. So we get $(\sqrt{x}-\sqrt{x}_+) \to \abs{\sqrt{x}-\sqrt{x}_+}e^{-i \pi}=(\sqrt{x}_+-\sqrt{x})e^{-i \pi}$. Therefore, the outgoing mode at the outer horizon surface becomes
\begin{equation}
\Phi_{\mathrm{out}}(x<x_+)= e^{-i\w v}(\sqrt{x}_+-\sqrt{x})^{i\w/\k_+}e^{\pi \omega / \k_+}. \label{analyticmode}
\end{equation}
Thus \eqref{nonanalyticmode} and \eqref{analyticmode} describe the outgoing mode outside and inside the outer horizon and we can use this to find out the relative scattering probability of the scalar field at the outer horizon,
\begin{equation}
    \Gamma_+ = \biggl|\dfrac{\Phi_{\mathrm{out}}(x>x_+)}{\Phi_{\mathrm{out}}(x<x_+)} \biggl|^2=e^{-2 \pi \frac{\w}{\k_+}}.
\end{equation}
We can now calculate the Hawking radiation spectrum which is given by
\begin{equation}
    |N_\w|^2=\dfrac{1}{e^{2 \pi {\w}/{\k_+}}-1}=\dfrac{1}{e^{\hbar \w /k_B T_+}-1},
\end{equation}
where $k_B T_+= \hbar \k_+ /2 \pi$, $T_+$ being the Hawking temperature.

At this juncture, we would have ideally preferred to investigate the internal structure of this black hole solution. However, we face an impasse: to study the internal structure, we have to construct solutions to the confluent Heun equation which satisfy the specified boundary conditions at two of the singular points simultaneously, but the so-called \textit{connection problem} has not yet been solved for the Heun class of differential equations. So we turn our attention to a black hole spacetime which is much more amenable to such calculations.
\section{The Dimensionally Reduced BTZ Black Hole}\label{aoblackhole}
\begin{figure}[h!]
	\centering
	\includegraphics[width=50mm]{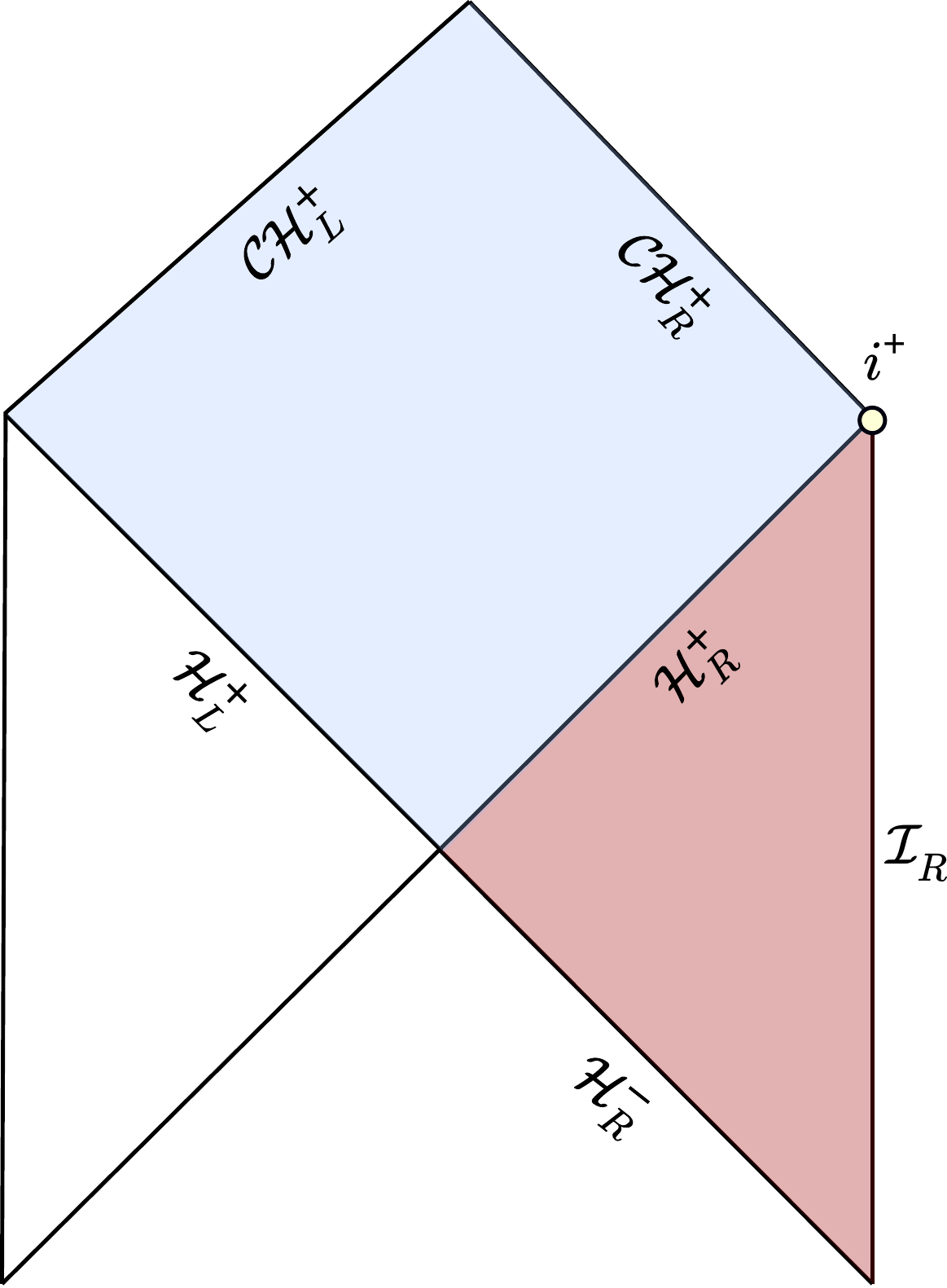}
	\caption{Penrose diagram of a dimensionally reduced BTZ black hole: the blue region denotes the interior of the black hole and the red region denotes the exterior. The future event horizon ($\mathcal{H}^+_R$), the past event horizon ($\mathcal{H}^-_R$) the right and left future Cauchy horizons ($\mathcal{CH}^+_R$ and $\mathcal{CH}^+_L$, respectively) have been indicated in the diagram. $\mathcal{H}^+_L$ is the future even horizon of the other asymptotically AdS region.}
	\label{aobh_FIGURE}
\end{figure}

Starting from Einstein gravity in three dimensions with a negative cosmological constant, one can perform a dimensional reduction to a model similar to Jackiw-Teitelboim gravity in two dimensions \cite{Achucarro:1993fd, NavarroSalas:1999up} . The action for the two-dimensional theory is given by
\begin{equation}
    S= \int \d^2 x \sqrt{-g} \phi \left(R-2 \Lambda - \dfrac{J}{2 \phi^4}\right),
\end{equation}
where $\Lambda=-1/L^2$, $L$ being the AdS radius, $J$ is a constant and corresponds to a charge in the two-dimensional theory and $\phi$ is the dilaton field. This theory has a black hole solution given by
\begin{equation}
ds^2= -f(r)dt^2+f(r)^{-1}dr^2,  ~ \phi=\dfrac{r}{L},\label{aometric}
\end{equation}
where
\begin{equation}
f(r)=-\Lambda r^2 -M + \dfrac{J^2}{4r^2} \label{aolapse}=\dfrac{(r^2-r_+^2)(r^2-r_-^2)}{L^2r^2}. 
\end{equation}
Here $r_+$ and $r_-$ are the outer and inner horizons and are given by
\begin{equation}
r_\pm=\dfrac{ML^2}{2}\left(1 \pm \sqrt{1-\left(\dfrac{J}{ML}\right)^2} \right).
\end{equation}
Since this black hole clearly corresponds the $t,r$ section of a BTZ black, they have an identical causal structure as shown in Fig. \ref{aobh_FIGURE}. We also note that we can define the ingoing and outgoing Eddington-Finkelstein coordinates, $v$ and $u$ respectively, along with a tortoise coordinate $r_*$ for this spacetime in the usual way.
\subsection{Scalar field perturbations at $\mathcal{CH}^{+}_{R}$}
To study scalar field perturbations in the spacetime described by \eqref{aometric} we again use the generalized form of the Klein-Gordon equation, that is, one with the coupling $h(\phi)$,
\begin{equation}
    \dfrac{1}{\sqrt{-g} h(\phi)}\partial_\mu \left( g^{\mu \nu}  \sqrt{-g} h(\phi)  \partial_\nu \Phi(r,t) \right) - {\mu^2}\Phi(r,t) =0.
\end{equation}
Previous attempts \cite{Cordero:2012je} at studying scalar perturbations of this spacetime revealed that the resultant radial equation was too complicated for exact calculations. Using the generalized Klein-Gordon equation however makes exact calculations possible. Following arguments we have already discussed, we set $h(\phi)=\phi$ and implement the gauge $\phi=r/L$ to simplify the above equation using the ansatz 
\begin{subequations} 
   \begin{align}
           \Phi(r,t) &=e^{-i \omega t}R(r) \label{phiansatz}, \\
           &\equiv e^{-i \omega t}z^{-i \frac{\omega}{2 \kappa_-}}(1-z)^{-i\frac{\omega}{2\kappa_+}}F(z),\label{ansatz}
   \end{align}
\end{subequations}
where we have used a new radial coordinate following \cite{Dias:2019ery},
\begin{equation}
    z=\dfrac{r^2-r^2_-}{r^2_+-r^2_-}.
\end{equation}
We find that $F(z)$ satisfies the hypergeometric equation
\begin{equation}
    z(1-z) \partial_z^2F(z)+[c-(a+b+1)z]\partial_z F(z)-abF(z)=0 \label{hypgeo}.
\end{equation}
where
\begin{align}
    \label{abc}
    & a=\dfrac{1}{2}\left( \Delta -i\dfrac{\omega}{\kappa_-} -i \dfrac{\omega}{\kappa_+} \right), \nonumber \\ &  b=\dfrac{1}{2}\left( 2- \Delta -i\dfrac{\omega}{\kappa_-} -i \dfrac{\omega}{\kappa_+} \right), \nonumber ~ c=1- i \dfrac{\omega}{\kappa_-}, \\ & \mu^2 L^2= \Delta(\Delta-2).
\end{align}
Here $\Delta$ is a parameter that encodes the mass $\mu$ of the scalar field and the boundary conditions that it satisfies \cite{Breitenlohner:1982jf}. We also recall that \eqref{hypgeo} has three singular points, namely $z=0,1,\infty$. We can write down a set of two linearly independent solutions around each singular point and each set forms a basis. The following discussion closely mirrors that of \cite{Dias:2019ery} given that the spacetime under consideration is merely the $t,r$ section of a rotating BTZ black hole. We therefore summarize their analysis while adapting it to our setting.

So, inside the black hole ($r_-<r<r_+, 0<z<1$), we can write down a basis for $R(z)$ using \eqref{ansatz} and the two linearly independent solutions to \eqref{hypgeo} \cite{Nikiforov1988} as
\begin{subequations} 
\label{CHbasis}
\begin{align}
     R_{\mathrm{out},-}= &z^{-\frac{1}{2}(1-c)}(1-z)^{\frac{1}{2}(a+b-c)}{_2F_1(a,b;c;z)}, \\
     R_{\mathrm{in},-}= & z^{\frac{1}{2}(1-c)}(1-z)^{\frac{1}{2}(a+b-c)} \nonumber \\ &\times {_2F_1(a-c+1,b-c+1;2-c;z)},
\end{align}
\end{subequations}
and at the Cauchy horizon ($z=0$), these two linearly independent solutions behave as
\begin{subequations}
\begin{align}
   & R_{\mathrm{out},-}|_{z \sim 0}=z^{-i\frac{\omega}{2\kappa_-}}\hat{R}_{\mathrm{out},-}(\omega;z), \\
   & R_{\mathrm{in},-}|_{z \sim 0}=z^{+i\frac{\omega}{2\kappa_-}}\hat{R}_{\mathrm{in},-}(\omega;z),
\end{align}
\end{subequations}
where $\hat{R}_{\mathrm{out},-}(\omega;z)$ and $\hat{R}_{\mathrm{in},-}(\omega;z)$ are analytic at $z=0$ and are equal to unity since $_2F_1(\alpha,\beta;\gamma;0)=1$. Moreover, for $z\neq0$, $_2F_1(\alpha,\beta;\gamma;0)=1$ has simple poles at $\gamma=-N$ (where $N=0,1,2, \cdots$): it is analytic otherwise. Therefore we can deduce that $R_{\mathrm{in},-}$ has simple poles when $\omega$ is a positive integer multiple of $ik_-$ and $R_{\mathrm{out},-}$ has simple poles when $\omega$ is a negative integer multiple of $ik_-$.  These two linearly independent basis solutions give rise to two sets of modes through \eqref{phiansatz} which are labeled as $\Phi_{\mathrm{out},-} (\propto R_{\mathrm{out},-})$ and $\Phi_{\mathrm{in},-}(\propto R_{\mathrm{in},-})$ and are called the outgoing and ingoing modes, respectively. Noting that the outgoing Eddington-Finkelstein coordinates $(u,r)$ are regular across the right Cauchy horizon ($\mathcal{CH}^+_R$), we can write the modes near $\mathcal{CH}^+_R$, as
\begin{subequations}
    \label{modeCHR}
\begin{align}
    & \Phi_{\mathrm{out},-} = e^{i\omega(u-u_0)}\left( 1 + \mathcal{O}(z)\right), \\
    & \Phi_{\mathrm{in},-} = e^{i\omega(u-u_0)}z^{+i\frac{\omega}{\kappa_-}}\left( 1 + \mathcal{O}(z)\right).
\end{align}
\end{subequations}
Similarly, since the ingoing Eddington-Finkelstein coordinates $(v,r)$ are regular across the left Cauchy  horizon ($\mathcal{CH}^+_L$), we can write the modes near $\mathcal{CH}^+_L$ as 
\begin{subequations}
    \label{modeCHL}
\begin{align}
    & \Phi_{\mathrm{out},-} = e^{i\omega(v-v_0)}z^{+i\frac{\omega}{\kappa_-}}\left( 1 + \mathcal{O}(z)\right), \\
    & \Phi_{\mathrm{in},-} = e^{i\omega(v-v_0)}\left( 1 + \mathcal{O}(z)\right).
\end{align}
\end{subequations}
Here, $u_0$ and $v_0$ are real constants that depend on the black hole parameters. More importantly, we can see from \eqref{modeCHR} and \eqref{modeCHL} that $\Phi_{\mathrm{out},-}$ is smooth across  $\mathcal{CH}^+_R$ but not across  $\mathcal{CH}^+_L$, whereas $\Phi_{\mathrm{in},-}$ is smooth across $\mathcal{CH}^+_L$ but not across $\mathcal{CH}^+_L$. Likewise, we can consider solutions $R_{\mathrm{out},+}$ and $R_{\mathrm{in},+}$ near the event horizon $z=1$ which can be written in a similar fashion near $r=r_+ (z=1)$, viz.,
\begin{subequations}
\label{ehbasis2}
\begin{align}
   & R_{\mathrm{out},+}|_{z \sim 1}=(1-z)^{+i\frac{\omega}{2\kappa_+}}\hat{R}_{\mathrm{out},+}(\omega;z), \\
   & R_{\mathrm{in},+}|_{z \sim 1}=(1-z)^{-i\frac{\omega}{2\kappa_+}}\hat{R}_{\mathrm{in},+}(\omega;z), 
\end{align}
\end{subequations}
where $\hat{R}_{\mathrm{out},+}(\omega;z)$ and $\hat{R}_{\mathrm{in},+}(\omega;z)$ are analytic at $z=1$ and we also note (arguing as before) that $R_{\mathrm{in},+}$ has simple poles when $\omega$ is negative integer multiple of $ik_+$ and $R_{\mathrm{out},+}$ has simple poles when $\omega$ is positive integer multiple of $ik_+$. These two linearly independent solutions give rise two sets of modes, $\Phi_{\mathrm{out},+}(\propto R_{\mathrm{out},+})$ and $\Phi_{\mathrm{in},+}(\propto R_{\mathrm{in},+})$. We can also show that $R_{\mathrm{in},+}$ is smooth across $\mathcal{H}^+_R$.
Similarly, we have the linearly independent solutions $R_{\mathrm{vev},\infty}$ and $R_{\mathrm{in},\infty}$ which near $\mathcal{I}_R$ behaves as
\begin{subequations}
\begin{align}
   & R_{\mathrm{vev},\infty}|_{z \sim \infty}= z^{-\Delta/2}\left( 1 + \mathcal{O}(1/z)\right), \label{vev}\\
   & R_{\mathrm{in},\infty}|_{z \sim \infty}= z^{-(2-\Delta)/2}\left( 1 + \mathcal{O}(1/z)\right), 
\end{align}
\end{subequations}
and give rise to the modes $\Phi_{\mathrm{vev},\infty}$ and $\Phi_{\mathrm{source},\infty}$ respectively.
Using the linear transformation relations for hypergeometric functions \cite{Nikiforov1988}, we can express the event horizon basis  in terms of the Cauchy horizon basis,
\begin{subequations}
\label{ehtoCHbasis}
\begin{align}
    & R_{\mathrm{out},+}= \mathcal{A}(\omega) R_{\mathrm{out},-}+\mathcal{B}(\omega) R_{\mathrm{in},-}, \label{outscat}\\
    & R_{\mathrm{in},+}= \tilde{\mathcal{A}}(\omega) R_{\mathrm{in},-}+\tilde{\mathcal{B}}(\omega) R_{\mathrm{out},-}, \label{inscat}
\end{align}
\end{subequations}
and we can also write $R_{\mathrm{in},+}$ in terms of the $\mathcal{I}_R$ basis, and $R_{\mathrm{vev},\infty}$ in terms of the event horizon basis
\begin{subequations}
\begin{align}
    & R_{\mathrm{in},+}= \dfrac{1}{\mathcal{T}(\omega)} R_{\mathrm{source},\infty}+\dfrac{\mathcal{R}(\omega)}{\mathcal{T}(\omega)} R_{\mathrm{vev},\infty}, \label{rin2inf}\\
    & R_{\mathrm{vev},\infty}= \dfrac{1}{\tilde{\mathcal{T}}(\omega)} R_{\mathrm{out},+}+\dfrac{\tilde{\mathcal{R}}(\omega)}{{\tilde{\mathcal{T}}(\omega)}} R_{\mathrm{in},+}. \label{vev2eh}
\end{align}
\end{subequations}
In \eqref{ehtoCHbasis}, $\mathcal{A}$ and $\mathcal{B}$ represent the transmission and reflection coefficients for scattering of waves propagating out from $\mathcal{H}_L^+$, and $\tilde{\mathcal{A}}$ and $\tilde{\mathcal{B}}$ represent the transmission and reflection coefficients for scattering of waves propagating in from $\mathcal{H}_R^+$; in \eqref{rin2inf}, $\mathcal{T}$ and $\mathcal{R}$ represent the transmission and reflection coefficients for scattering of waves incident from $\mathcal{I}_R$; in \eqref{vev2eh} $\tilde{\mathcal{T}}$ and $\tilde{\mathcal{R}}$ represent the transmission and reflection coefficients for scattering of waves propagating out of $\mathcal{H}_R^-$ \cite{Dias:2018etb}. The explicit expressions of these coefficients are given below,
\begin{subequations}
\label{gammafns}
\begin{align}
&\mathcal{A}(\omega)=\dfrac{\Gamma{(1-c)}\Gamma{(1-a-b+c)}}{\Gamma{(1-a)}\Gamma{(1-b)}}, \nonumber \\ &\mathcal{B}(\omega)=\dfrac{\Gamma{(c-1)}\Gamma{(-a-b+c+1)}}{\Gamma{(c-a)}\Gamma{(c-b)}}, \\
&\tilde{\mathcal{A}}(\omega)=\dfrac{\Gamma{(c-1)}\Gamma{(a+b-c+1)}}{\Gamma{(a)}\Gamma{(b)}}, \nonumber \\ &\tilde{\mathcal{B}}(\omega)=\dfrac{\Gamma{(1-c)}\Gamma{(a+b-c+1)}}{\Gamma{(a-c+1)}\Gamma{(b-c+1)}}, \\
&\mathcal{T}(\omega)=\dfrac{\Gamma{(a)}\Gamma{(a-c+1)}}{\Gamma{(a-b)}\Gamma{(a+b-c+1)}}, \label{TR} \nonumber \\ &\mathcal{R}(\omega)=\dfrac{\Gamma{(a)}\Gamma{(b-a)}\Gamma{(a-c+1)}}{\Gamma{(b)}\Gamma{(a-b)}\Gamma{(b-c+1)}}, \\
&\tilde{\mathcal{T}}(\omega)=\dfrac{\Gamma{(a)}\Gamma{(a-c+1)}}{\Gamma{(a-b+1)}\Gamma{(a+b-c)}}, \label{TtRt} \nonumber \\ &\tilde{\mathcal{R}}(\omega)=\dfrac{\Gamma{(a)}\Gamma{(a-c+1)}\Gamma{(-a-b+c)}}{\Gamma{(1-b)}\Gamma{(c-b)}\Gamma{(a+b-c)}}.
\end{align}
\end{subequations}

\subsubsection{Exterior quasinormal modes}
The linear mode solutions which satisfy the ``no-source" boundary condition at $\mathcal{I}_R$ and are smooth at the event horizon $\mathcal{H}_R^+$ are known as the exterior quasinormal modes \cite{Dias:2019ery}. In literature, the frequencies of these modes are referred just as quasinormal modes (QNMs) \cite{Kokkotas:1999bd,Berti:2009kk,Konoplya:2011qq}.

The no-source boundary condition actually translates to a vanishing boundary condition at $\mathcal{I}_R$. This means that the radial function is strictly proportional to $R_{\mathrm{vev},\infty}$ at $\mathcal{I}_R$ and as we can see from \eqref{vev}, $R_{\mathrm{vev},\infty} \to 0$ as $r \to \infty$. Moreover, smoothness across the event horizon $\mathcal{H}^+_R$ would imply that the radial function is proportional to $R_{\mathrm{in},+}$ \cite{Dias:2019ery}. This gives us the defining condition of a quasinormal mode, viz $R_{\mathrm{in},+} \propto R_{\mathrm{vev},\infty}$. This is equivalent to setting $\mathcal{T}(\omega)=\infty$ or $\tilde{\mathcal{T}}(\omega)=\infty$ in \eqref{rin2inf} and \eqref{vev2eh}, respectively. From the properties of the gamma functions, this translates to setting $a=-n$ or $a-c+1=-n$ for $n=0,1,2, \cdots$ in \eqref{TR} or \eqref{TtRt}. Using \eqref{abc}, these gives us two sets of QNMs which we called the left and right quasinormal modes, denoted $\omega_L$ and $\omega_R$, where
\begin{subequations}
\begin{align}
    & \omega_L=-i\dfrac{r_+-r_-}{L^2}(\Delta +2n), \label{omegaL}\\
    & \omega_R=-i\dfrac{r_+ + r_-}{L^2}(\Delta + 2n), \label{omegaR} 
\end{align}
\end{subequations}
for $n=0,1,2,\cdots$. We note that the quasinormal mode frequencies are purely imaginary.

\subsubsection{Interior quasinormal modes}
The interior quasinormal modes are the zeros of the scattering amplitudes (reflection and transmission coefficients) inside the black hole. Let us consider \eqref{ehtoCHbasis} where we have written the event horizon solutions in the Cauchy horizon basis: \eqref{inscat} describes an ingoing wave, proportional to $\Phi_{\mathrm{in},+}$, coming in through the event horizon $\mathcal{H}^+_R$ and propagating to the Cauchy Horizon $\mathcal{CH}^+_{L,R}$. This is called ``in-scattering." Similarly \eqref{outscat} describes ``out-scattering," that is, a wave proportional to $\Phi_{\mathrm{out},+}$ coming out from $\mathcal{H}_L^+$ and propagating to the Cauchy horizon $\mathcal{CH}^+_{L,R}$. We can now naturally define the interior QNMs as follows: i) in-out interior QNMs [$\tilde{\mathcal{A}}(\omega)=0$] and in-in interior QNMs [$\tilde{\mathcal{B}}(\omega)=0$], and ii) out-in interior QNMs [${\mathcal{A}}(\omega)=0$] and out-out interior QNMs [${\mathcal{A}}(\omega)=0$] \cite{Dias:2019ery}.

The first pair is associated with \eqref{inscat} and the second pair with \eqref{outscat}. We shall focus here on the first pair of QNMs.
The in-out interior QNMs describes modes that enter from the event horizon $\mathcal{H}^+_R$ and are completely reflected towards the right Cauchy horizon ($\mathcal{CH}_R^+$). Since this indicates $\tilde{\mathcal{A}}(\omega)=0$, we can set $a=-n$ or $b=-n$ for $n=0,1,2,\cdots$ and determine the spectrum of the in-out interior QNMs, viz,
\begin{subequations}
\begin{align}
    & \omega_{\mathrm{in-out,1}}=-i\dfrac{r_+-r_-}{L^2}(\Delta +2n), \label{omgeainout1}\\
    & \omega_{\mathrm{in-out,2}}=-i\dfrac{r_+ - r_-}{L^2}(2-\Delta + 2n). \label{omgeainout1}
\end{align}
\end{subequations}
We note that one of the in-out interior quasinormal mode frequencies is identical to the exterior left quasinormal mode frequency, that is
\begin{equation}
    \omega_{\mathrm{in-out,1}}=\omega_L.
\end{equation}

Similarly the in-in interior QNMs describe modes that come in from $\mathcal{H}_R^+$ and are completely transmitted towards the left Cauchy horizon ($\mathcal{CH}_L^+$). We can also show that one of the in-in QNM frequencies coincide with the exterior right quasinormal mode frequencies. We note in passing that the out-in interior QNMs describe modes that come out from $\mathcal{H}^+_L$ and are completely reflected inwards to $\mathcal{CH}_L^+$, and the out-out interior QNMs describe modes that come out from $\mathcal{H}^+_L$ and are completely transmitted outwards to $\mathcal{CH}_R^+$.

\subsubsection{Inner horizon instability and strong cosmic censorship} \label{aoihinstability}
To study the stability of the inner horizon, we may set up an initial value problem in the following manner \cite{Dias:2019ery}: we define smooth outgoing wave packets on $\mathcal{H}_L^+$ and $\mathcal{H}_R^-$ and turn on a source at $\mathcal{I}_R$ with profiles $Z(\w)$, $X(\w)$ and $\tilde{X(\w)}$ respectively. These initial boundary conditions uniquely determine a solution $\Phi$ inside and outside the black hole. Inside the black hole we can then write $\Phi$ as
\begin{equation}
\label{inoutdecomp}
    \Phi(x)= \Phi_{\mathrm{out}} (x)+ \Phi_{\mathrm{in}} (x),
\end{equation}
where 
\begin{subequations}
\begin{align}
	  \Phi_{\mathrm{out}}(x) \equiv & \int \d \omega \left( Z(\omega) \mathcal{A} (\omega) + \left(\tilde{X}(\w)\mathcal{T}(\w) \right. \right. \nonumber \\ & \left. \left. + X(\w)\tilde{\mathcal{R}}(\w)\right) \tilde{\mathcal{B}}(\omega)\right) \Phi_{\mathrm{out},-}(\omega), 
   \\
   \Phi_{\mathrm{in}}(x) \equiv & \int \d \omega \left( Z(\omega) \mathcal{B} (\omega) + \left(\tilde{X}(\w)\mathcal{T}(\w)\right. \right. \nonumber \\ & \left. \left.+X(\w)\tilde{\mathcal{R}}(\w)\right)(\omega) \tilde{\mathcal{A}}(\omega)\right) \Phi_{\mathrm{in},-}(\omega). 
\end{align}
\end{subequations}

We note that $\Phi_{\mathrm{out}}$ is smooth at $\mathcal{CH}^+_R$ and $\Phi_{\mathrm{in}}$ is smooth at $\mathcal{CH}^+_L$. Due care needs to be taken when splitting $\Phi$ into these two parts as argued in \cite{Dias:2019ery}. We refer the reader to Dias \textit{et al.} \cite{Dias:2019ery} for more technical details. 

Now, we look closely at the behavior of $\Phi$ at $\mathcal{CH}^+_R$: any nonsmooth behavior of $\Phi$ must come from $\Phi_{\mathrm{in}}$ since $\Phi_{\mathrm{out}}$ is smooth at $\mathcal{CH}^+_R$. Using \eqref{modeCHR} we can write $\Phi_{\mathrm{in}}$ as 
\begin{equation}
    \Phi_{\mathrm{in}}= \int_{C_0} \d \omega \mathcal{G}(\omega) e^{-i \omega (u-u_0)} \exp{\left[i\dfrac{\omega}{\kappa_-}\log z\right]}(1+\mathcal{O}(z)).
\end{equation}
where $C_0$ is a contour that is indented in such a way that it passes just below $\w=0$ in the complex $\w$ plane and
\begin{align}
    \mathcal{G} =   Z(\w) \mathcal{B}(\w) + \left( \tilde{X}(\w)\mathcal{T}(\w)+X(\w)\tilde{\mathcal{R}}(\w)\right)\tilde{\mathcal{A}(\w)}.
\end{align} 

We also note that the pole at $\w=0$ arises from the manner in which one chooses the basis to construct the solution and hence the contour of integration can be made to pass just below it \cite{Dias:2019ery}. The right Cauchy horizon is located at $z=0$ where $\log z \to - \infty$. The contour of integration to determine $\Phi_\mathrm{in}$  is deformed to a line of constant Im($\w$) in the lower half of the complex $\w$ plane. The contribution from the poles that are crossed with the deformed contour plays crucial role in the behavior of $\Phi_\mathrm{in}$ since the smoothness of $\mathcal{G}(\w)$ ultimately determines the regularity of $\Phi_\mathrm{in}$ near $\mathcal{CH}^+_R$. 

We begin by studying the expressions of $\tilde{\mathcal{A}}(\w)$, $\mathcal{B}(\w)$, $\mathcal{T}(\w)$ and $\tilde{\mathcal{R}}(\w)$: putting \eqref{abc} in \eqref{gammafns} we get
\begin{subequations}
\begin{align}
    \tilde{\mathcal{A}}(\w) &=\dfrac{i \k_-}{\w}\dfrac{\Gamma\left(1-i\frac{\w}{\k_-}\right)\Gamma \left(1-i\frac{\w}{\k_+}\right)}{\Gamma\left(\frac{\Delta}{2}+i\frac{L}{2}\frac{\w L}{r_--r_+}\right)\Gamma\left(1-\frac{\Delta}{2}+i\frac{L}{2}\frac{\w L}{r_--r_+}\right)}, \\
     \tilde{\mathcal{B}}(\w) &=\dfrac{i \k_-}{\w}\dfrac{\Gamma\left(1-i\frac{\w}{\k_-}\right)\Gamma \left(1+i\frac{\w}{\k_+}\right)}{\Gamma\left(\frac{\Delta}{2}+i\frac{L}{2}\frac{\w L}{r_-+r_+}\right)\Gamma\left(1-\frac{\Delta}{2}+i\frac{L}{2}\frac{\w L}{r_-+r_+}\right)}, \\
     \mathcal{T}(\w) &=\dfrac{\Gamma\left( \frac{\Delta}{2}-i\frac{L}{2}\frac{\w L}{r_+ - r_-}\right)\Gamma\left( \frac{\Delta}{2}-i\frac{L}{2}\frac{\omega L}{r_+ + r_-} \right)}{\Gamma\left( 1-i\frac{\w}{\k_+}\right)\Gamma\left( \Delta -1\right)}, \\
      \tilde{\mathcal{R}}(\w) &=\dfrac{\Gamma\left(1+i\frac{\w}{\k_+}\right)\Gamma\left( \frac{\Delta}{2}-i\frac{L}{2}\frac{\w L}{r_+ - r_-}\right)\Gamma\left( \frac{\Delta}{2}-i\frac{L}{2}\frac{\omega L}{r_+ + r_-} \right)}{\Gamma\left(1-i\frac{\w}{\k_+}\right)\Gamma\left( \frac{\Delta}{2}+i\frac{L}{2}\frac{\w L}{r_+ - r_-}\right)\Gamma\left( \frac{\Delta}{2}+i\frac{L}{2}\frac{\omega L}{r_+ + r_-} \right)}.
\end{align}
\end{subequations}
Consider the expressions for $\tilde{\mathcal{A}}(\w)$ and $\mathcal{B}(\w)$: both of these scattering coefficients have simple poles at $\w=0$. From the first gamma function in the numerator, they also have simple poles at $\w=-in\k_-$ for $n=1,2,3, \cdots$. Moreover, from the second gamma function in each numerator, we can see that $\tilde{\mathcal{A}}(\w)$ has simple poles at $\w=-in\k_+$, and $\mathcal{B}(\w)$ has simple poles at $\w=+in\k_+$.
$\mathcal{T}(\w)$ has simple poles at the exterior quasinormal mode frequencies $\w=\w_{L,R}$ (this, as we have seen, is the definition of exterior QNMs). $\tilde{\mathcal{R}}(\w)$ also has simple poles at $\w=\w_{L,R}$ and additional simple poles at $\omega=+in\k_+$ in the upper half plane. We also note that both $\mathcal{T}(\w)$ and $\tilde{\mathcal{R}}(\w)$ have zeros  at $\w=-in\k_+$ which arise from the first gamma function in their respective denominators. 

With this information, we can determine the analyticity of $\mathcal{G}(\w)$. We are interested in the lower half plane because that is where we deform our contour of integration. Since $Z(\w)$, $X(\w)$ and $\tilde{X}(\w)$ actually correspond to Fourier transforms of compactly supported initial data, they entire functions of $\w$. So the singularities in $\mathcal{G}(\w)$ arise from the singularities in the scattering coefficients that we have just mentioned, namely, it arises from $\mathcal{B}$, $\mathcal{T}\tilde{\mathcal{A}}$ and $\mathcal{R}\tilde{\mathcal{A}}$. We discuss the nature of each of these three objects below:

$\mathcal{B}$ has a pole at $\w=0$ and we have chosen the contour of integration to pass below this pole. Hence it does not affect the freedom to deform the contour of integration into the lower half plane. The only poles of $\mathcal{B}$ in the lower half plane lie at $\w=-in\k_-$. 

 $\mathcal{T}\tilde{\mathcal{A}}$ has the following pole structure: $\tilde{\mathcal{A}}$ has a pole at $\w=0$ which is irrelevant. It also has poles  at $\w=-in\k_\pm$. However $\mathcal{T}$ has a zero at $\w=-ink_+$. So only the poles at $\w=-in\k_-$ from $\tilde{\mathcal{A}}$ contribute to the product  $\mathcal{T}\tilde{\mathcal{A}}$. Now  $\mathcal{T}$ has poles at $\w=\w_{L,R}$. But we have seen from the discussion on interior QNMs, the ``in-out" QNM $\w_{\mathrm{in-out},1}=w_L$ is a zero of $\tilde{\mathcal{A}}$. So the pole at $\w=\w_L$ gets canceled but no such cancellation occurs for $\w=\w_R$. So we see that $\mathcal{T}\tilde{\mathcal{A}}$ has two poles in the lower half plane, namely, at $\w=\w_R$ and $\w=in\k_-$ for $n=1,2,3, \cdots$. 

\begin{figure}[h!]
	\centering
	\includegraphics[width=85mm]{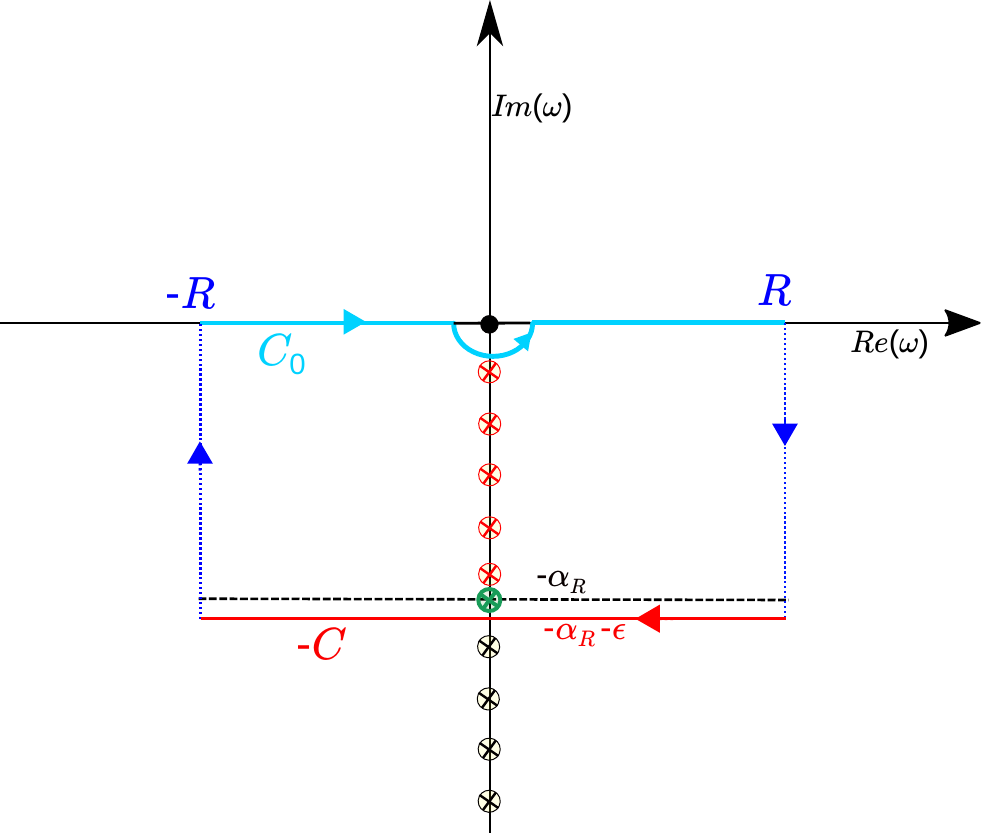}
 	\caption{Schematic diagram of the contours of integration in the complex $\w$ plane and the relevant poles. The pole marked in green is the slowest decaying right quasinormal mode. The remaining poles correspond to $\w=-in\k_-$, the ones marked in red contribute (smoothly) to the integral and the black ones do not.}
	\label{pole_FIGURE}
\end{figure}

By a very similar argument we can show that $\mathcal{R}\tilde{\mathcal{A}}$ has a pole structure identical to $\mathcal{T}\tilde{\mathcal{A}}$ in the lower half plane. Therefore, for compactly supported initial data, $\mathcal{G}$ is analytic in the lower half plane except for simple poles at $\w=\w_R$ and $\w=-in\k_-$ (see Fig \ref{pole_FIGURE}). 

Let $\varpi$ be the frequency of the slowest decaying right quasinormal mode, then we have 
\begin{equation}
    \varpi=-i\dfrac{(r_+ +r_-)\Delta}{L^2} \implies \alpha_R \equiv-\mathrm{Im}(\varpi)=\dfrac{(r_++r_-)\Delta}{L^2},
\end{equation}
where we have define the ``right spectral gap", $\alpha_R$.
We now deform the contour of integration $C_0$ into a new contour $C$ defined as the straight line Im($\w$)=$-\alpha_R-\epsilon$, that is, we push the contour just beyond the pole at $\varpi_m$. In doing so, we pick up a contribution from the pole at $\varpi$. We also pick up from the poles at $\w=-in\k_-$ which lie between $C_0$ and $C$ (see Fig \ref{pole_FIGURE}). But by applying the residue theorem, we see that the poles at  at $\w=-in\k_-$ gives a term which goes as $z^n$. Therefore these poles give a contribution which vanishes smoothly at $z=0$, that is, at the Cauchy horizon ($\mathcal{CH}_R^+$).
So the nonsmooth part of $\Phi_{in}$ arises from the poles at $\varpi$ and from the integration along $C$. Using the residue theorem, we can write
\begin{equation}
    -2 \pi i z^\beta \mathcal{G}_{\varpi}e^{-i \varpi (u-u_0)}(1+\mathcal{O}(z)),
\end{equation}
where $\mathcal{G}_{\varpi}$ is the residue of $\mathcal{G}$ at $\w=\varpi$ and $\beta=\alpha_R/\k_-$.

From this expression we see that the gradient of the scalar field will diverge at $\mathcal{CH}_R^+$ if $\beta < 1$. Thus $\beta<1$ ensures the energy momentum tensor of the scalar field will diverge at the Cauchy horizon as 
\begin{equation}
T_{VV}\sim V^{2(\beta-1)},
\label{ST}\end{equation} and potentially render it singular. When the black hole is far from extremality, $\beta<1$ and hence strong cosmic censorship will be respected. For near extremal black holes, $\beta>1$, therefore the stress energy tensor will be bounded at the right Cauchy horizon and such a black hole will violate strong cosmic censorship. This mimics the situation depicted in \cite{Dias:2019ery,Bhattacharjee:2020gbo}. 
\subsection{Shock wave singularity at $\mathcal{CH}^{+}_{L}$}
In this subsection, following \cite{Marolf:2011dj}, we study the singularity structure of $\mathcal{CH}^{+}_{L}$. We consider the late time geodesics falling into the black hole. These geodesics are characterized by the energy $E \equiv -u_t >0$ of the infalling observer and they satisfy the radial equation
\begin{equation}
    \dfrac{\d r}{d \tau}=-\sqrt{E^2-f(r)}.
\end{equation}
Here $\tau$ denotes the proper time and $u^\alpha$ denotes the four-velocity of the infalling observer. Near the inner horizon $r=r_-$, the above equation can be approximated as
\begin{equation}
\label{marolforiE}
    \dfrac{\d r}{d \tau} \approx -E.
\end{equation}
We now write the metric \eqref{aometric} in terms of double null coordinates $u=t-r_*$ and $v=t+r_*$, that is, $\d s^2= -f(r) \d u \d v$. We note that $u$ is past directed in the region $r_-<r<r_+$. Moreover, since $r_* \to \infty$ at $r=r_-$, $u$ or $v$ must diverge there. As we have mentioned, this means that $v$ is regular at $\mathcal{CH}^{+}_{L}$ and $u$ diverges to $-\infty$. We also approximate $f(r)$ near $r_-$ as $f(r)\approx -2 \k_- (r-r_-)$, since $\k_-\equiv (1/2)|\d f/\d r|_{r=r_-}$. We can then use the definition of the tortoise coordinate $\d r/\d r_*=f(r)$ to write
\begin{equation}
\label{MAROLFORI_R}
    r-r_-=Me^{2 \k_- r_*},
\end{equation}
where $M$ is taken as convenient prefactor to fix the integration constant in the definition of $r_*$.

In the previous section, we have mentioned that inside the black hole, we can decompose the scalar perturbation into an ingoing mode and an outgoing mode \eqref{inoutdecomp}, and we can further recast in the following form near $r=r_-$:
\begin{align}
    \Phi &= \Phi_{\mathrm{out}}(u)+ \Phi_{\mathrm{in}}(v)\nonumber \\ &= \int \d \w F(\w) e^{-i \w (u-u_0)}+ \int \d \w G(\w) e^{-i \w (v-v_0)},
\end{align}
where the exact expressions of $F(\w)$ and $G(\w)$ are inferred from the previous section (also see \cite{Balasubramanian:2004zu}). The important thing to note here is $\Phi_{\mathrm{in}}$ is smooth at $\mathcal{CH}^{+}_{L}$, whereas $\Phi_{\mathrm{out}}$ grows monotonically and blows up at $\mathcal{CH}^{+}_{L}$. So near the left Cauchy horizon ($\mathcal{CH}^{+}_{L}$) we can approximate $\Phi \approx \Phi_{\mathrm{out}}(u)$.

Let us now explore the behavior of $\Phi$ along the worldline of the infalling observer as a function of proper time $\tau$. We shall focus on late-time observers, this means that the value of $v=v_+$ when the observer crosses the event horizon $r_+$ is very large. We assume that $v_+ \gg M$. We also assume that observer crosses $\mathcal{CH}^{+}_{L}$ at $v_-$ such that $v$ increases monotonically along a timelike worldline. Moreover due to the time translation symmetry of the metric, $\Delta v= v_- - v_+=\Delta v(E)$ \cite{Marolf:2011dj}.

Now we shall estimate the proper times $\tau_{1,2}$ at the two events where the worldline intersect the null lines (say) $u=u_{1,2}$. From the assumption $v_+ \gg M$, it follows that $r_* \gg M$ and hence $r \approx r_-$ \cite{Marolf:2011dj}. So we can use \eqref{marolforiE} and \eqref{MAROLFORI_R} to write
\begin{equation}
    \tau \approx -\dfrac{r-r_-}{E} = -\dfrac{M}{E} e^{-2 \kappa_- r_*},
\end{equation}
where we have set $\tau=0$ at the worldline's intersection with $r_-$. Using $r_*=(v-u)/2$, we can write for $u=u_{1,2}$
\begin{equation}
    \tau_{1,2} \approx -\dfrac{M}{E}e^{\kappa_-(u_{1,2}-v_{1,2})},
\end{equation}
where $v_{1,2}$ denote the values of $v$ at the intersection of the worldline with $u_{1,2}$. Since $v_{1,2}>v_+$, we can readily calculate the upper bound
\begin{equation}
    |\tau_{1,2}|< \left(\dfrac{M}{E}e^{\k_- u_{1,2}} \right)e^{-\k_- v_+},
\end{equation}
and therefore the time interval $\Delta \tau=\tau_2-\tau_1>0$ is bounded by
\begin{equation}
    \Delta \tau < \left(\dfrac{M}{E}e^{\k_- u_{1}} \right)e^{-\k_- v_+}.
\end{equation}
The above equation tells us that, given the monotonically increasing behavior of $\Phi(u)$ near  $\mathcal{CH}^{+}_{L}$, the infalling observer will see the profile of the scalar field rise to a very high value within an arbitrarily short proper time interval that is proportional to $\exp(-\kappa_- v_+)$. For a sufficiently large value of $v_+$, the scalar perturbation will appear as a sharp shockwave of finite amplitude. This is a generic feature of black hole spacetimes \cite{Pandya:2020ejc} and was described first in \cite{Marolf:2011dj}. The observer will detect the shock wave effectively at $\tau=0$, that is, in the limit $v_+ \to \infty$, so it will be located just at the outgoing section of the inner horizon, that is, at $\mathcal{CH}^{+}_{L}$.
\subsection{Quantum effects at $\mathcal{CH}^{+}_{R}$} \label{quantumeffects}
\begin{figure}[htbp]
	\centering
	\includegraphics[width=60mm]{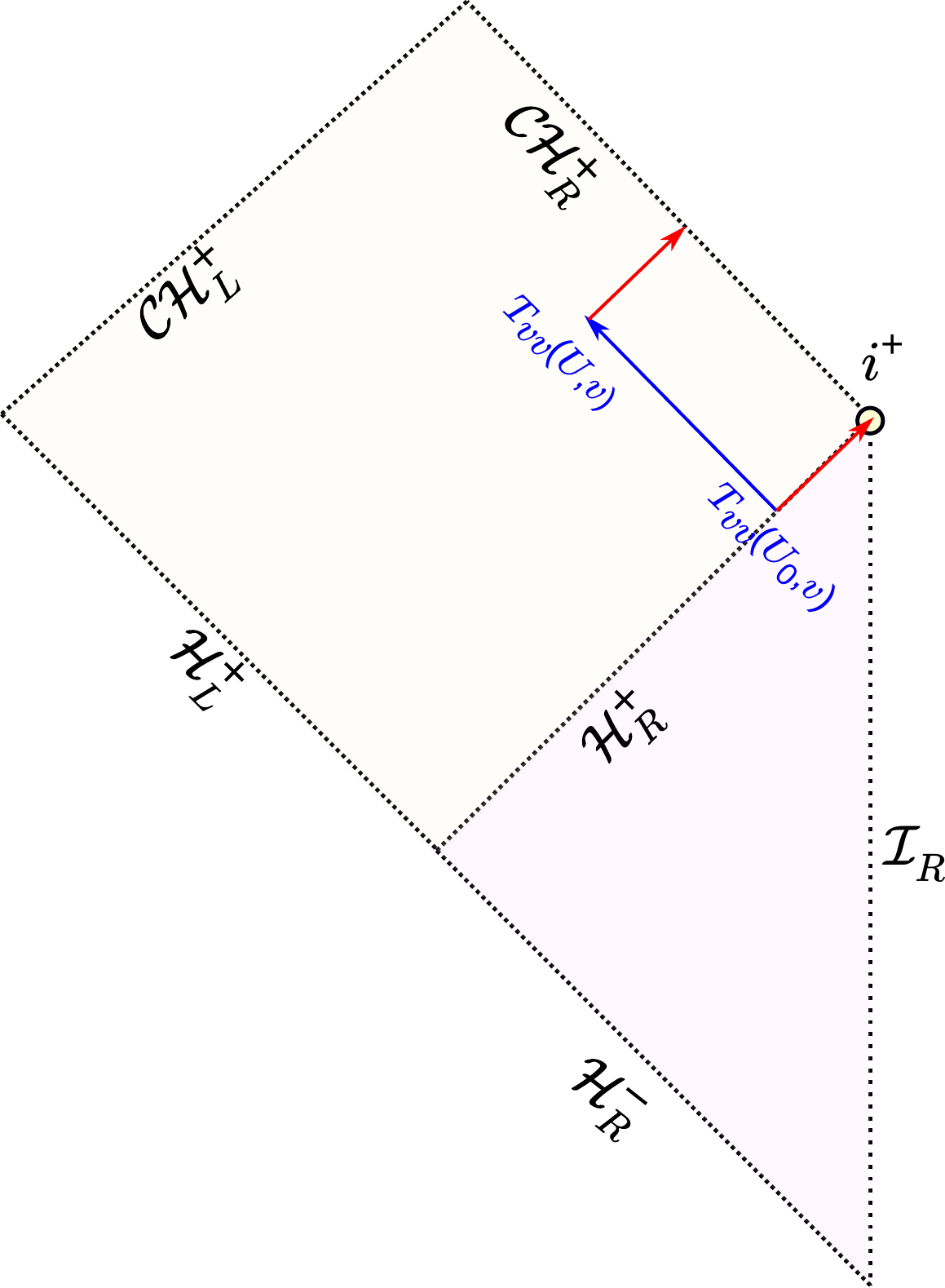}
	\caption{The blue arrow denotes the integration path and the red arrows show the limit $v \to \infty$ as discussed while calculating the quantum stress energy tensor in two dimensions.}
	\label{btz_quantum_FIGURE}
\end{figure}

After gaining reasonable insight on the possibility of violation of strong cosmic censorship for black holes in JT gravity, a pertinent question would be whether such a violation is seen if quantum effects are taken into account. This can however be easily analyzed using the trace anomaly of 2D quantum stress energy tensor of a probe massless scalar field as demonstrated in \cite{Birrell:1978th,Hollands:2019whz}. The analysis is quite similar to that of a 2D RN de Sitter black hole. To see this, we rewrite the solution (\ref{aometric}) again in terms of $u,v$ coordinates,
\begin{equation}
\d s^2=-f(r) \d u \d v.
\end{equation}
These double null coordinates $u, v$ are related to the Kruskal like coordinates $U, V$  that can be installed near $\mathcal{CH}^{+}_{R}$ via
\begin{equation}
   U=-e^{-\kappa_+u},\,\,V=-e^{-\kappa_- v}. \label{jtdnull}
\end{equation}
Similar coordinates can also be installed outside the event horizon, by changing the sign in front of $V$ to positive and $-\k_-$ to $+\k_+$ \cite{Dias:2019ery}. Now for any conformally invariant theory in two dimensions, the trace anomaly has a universal form (for a massless scalar field), 
\begin{equation}
    T^{\mu}_{\mu}=\dfrac{1}{24\pi}R,\label{tanmly}
\end{equation}
where $R$ is the Ricci scalar of the spacetime. For the metric (\ref{jtdnull}) $R=-f''(r)$\footnote{prime denotes derivative with respect to $r$.}, and using the $v$ component of continuity equation $\nabla_{\mu}T^{\mu\nu}=0$ we get, 
\begin{equation}
    \partial_uT_{vv}=-f\partial_v[f^{-1}T_{uv}].\end{equation}
    Then upon integrating, we get
  \begin{equation}  T_{Uv}=\dfrac{1}{192\pi}[2ff''-(f')^2]^{r(U,v)}_{r(U_0,v)}\,+T_{vv}(U_0,v), \label{toingrate}
\end{equation}
where the points $(U_0,v)$ and $(U,v)$ are situated on and inside the event horizon respectively (actually $U_0=0$ on the event horizon), along a null line  segment (see Fig \ref{btz_quantum_FIGURE}). Next, setting $v\to \infty$, the second term of (\ref{toingrate})  vanishes, and transforming into regular $V$ coordinate the first term gives
\begin{equation}
    T_{VV}\sim \dfrac{\kappa_+^2-\kappa_-^2}{\kappa_-^2}\dfrac{1}{V^2}.
\label{tvvdiv}\end{equation}
 This shows the quantum stress energy tensor diverges as one approaches ($V\to 0$) the right Cauchy horizon whenever $\kappa_+\neq\kappa_-$. Clearly, this divergence is stronger than the classical divergence obtained in (\ref{ST}) for $\beta <1$.  
\section{Discussion}
The paper has three distinct results. First, for a pure JT model, we have determined an exact analytical expression of scalar QNMs. Further, we have cross-checked this result numerically using the Horowitz-Hubeny prescription. Next, we have considered the 2D RN black hole and the massless scalar wave equation turned out to be the confluent Heun equation. Using the exact solution of the wave equation, we have employed the Damour-Ruffini method to find the Hawking spectra of the spacetime. We have also used the delta-condition to find some imaginary frequencies called resonant frequencies (in the Appendix). Obtaining the quasinormal frequencies would require numerical techniques, we wish to report the same in a future communication. In order to examine the strong cosmic censorship conjecture, we need the interior QNMs as well. In absence of those for the 2D RN black hole, we turned our attention towards the 2D analogue of rotating BTZ black hole. The dynamics of scalar field turned out to be quite similar to that of the usual BTZ black hole near the right Cauchy horizon. In fact the QNMs are just the same (after setting the azimuthal number $m$ to zero). We see that due to the coincidence where one of the interior QNMs matches with one of the exterior QNMs, the strong cosmic censorship conjecture is violated in the near extremal limit. The instability of the left Cauchy horizon is also depicted via the Marolf-Ori shock wave configuration. Finally, the quantum instability of right Cauchy horizon is shown to be more violent than the classical instability. 

This work offers obvious extensions. The interior QNMs for 2D RN black hole will shed some light on the interior structure of such spacetimes. We would like to attempt this in near future. Investigating other types of perturbations like vector and Weyl perturbations around these backgrounds would be useful to decipher whether the stability structure of these black holes change. It will also be interesting to see how the use of nonsmooth initial data changes the scenario. Investigating the backreaction of the Hawking quanta on background geometry and its influence on QNMs in lower dimensional models would be an interesting goal \cite{Martinez:1996gn, Konoplya:2004ik}. It would also be worthwhile to study scalar perturbations of black holes in deformed JT gravity \cite{Witten:2020wvy}. It will be interesting to try and understand the implications of these results in the context of the duality between the SYK model and the infrared sector of JT gravity \cite{Sachdev:1992fk,kitaev:talk,Maldacena:2016hyu}.
\section{Acknowledgments} \label{acknowledgment}
The research of S.B. and S.S. is supported by DST-SERB, Government of India, under the scheme Early Career Research Award (File no.: ECR/2017/002124) through the project titled \emph{``Near Horizon Structure of Black Holes."} The authors would like to also acknowledge support from IIIT, Allahabad through the Seed Grant for the project \emph{``Probing the Interior of $AdS$ Black Holes."} A.B. is supported by Research Initiation Grant (RIG/0300) provided by IIT, Gandhinagar and Start-Up Research Grant (SRG/2020/001380) by Department of Science and Technology, Science and Engineering Research Board (India). The computer algebra system \emph{Wolfram Mathematica} has been used to perform certain analytical and numerical calculations. 

\appendix*
\section{{Resonant frequencies of the 2D RN black hole}}

The standard confluent Heun function, $H_C$, reduces to a polynomial of degrees $N \geq 0$ if the following condition holds \cite{Fiziev_2009}:
\begin{equation}
    \dfrac{\delta}{\alpha}+\dfrac{\beta + \gamma}{2}+N+1=0.
\end{equation}
Using \eqref{tilde_params}, \eqref{alphabetagamma} and \eqref{etadelta} we can use the ``delta-condition" mentioned just above to get the following expression for the resonant frequencies $\w$ \cite{Vieira:2016ubt,Sakalli:2016fif}:
\begin{equation}
    \w = i \k_+ (1+N), ~ N=0,1,2,\cdots.
\end{equation}
In obtaining resonant frequencies, we have to impose boundary conditions on the solution such that it is finite at the horizon and well behaved at infinity. This necessitates the solution to have a polynomial form which is guaranteed by the form of $H_C(\alpha, \beta, \gamma, \delta, \eta,z)$ and the delta-condition. 

Now, before transforming to the confluent Heun equation, we could have also chosen the parameters with a minus and obtained modes which are apparently decaying.
It would be tempting to label these as the quasinormal frequencies. However, that would be premature since quasinormal frequencies correspond to modes that are purely ingoing at the event horizon and vanish at infinity. The delta-condition does not guarantee these boundary conditions. One needs to rigorously check this assertion numerically to reach a definite conclusion. In this regard, the approach used in \cite{Fiziev:2011mm} could be applicable. 


\begin{thebibliography}{63}%
	\makeatletter
	\providecommand \@ifxundefined [1]{%
		\@ifx{#1\undefined}
	}%
	\providecommand \@ifnum [1]{%
		\ifnum #1\expandafter \@firstoftwo
		\else \expandafter \@secondoftwo
		\fi
	}%
	\providecommand \@ifx [1]{%
		\ifx #1\expandafter \@firstoftwo
		\else \expandafter \@secondoftwo
		\fi
	}%
	\providecommand \natexlab [1]{#1}%
	\providecommand \enquote  [1]{``#1''}%
	\providecommand \bibnamefont  [1]{#1}%
	\providecommand \bibfnamefont [1]{#1}%
	\providecommand \citenamefont [1]{#1}%
	\providecommand \href@noop [0]{\@secondoftwo}%
	\providecommand \href [0]{\begingroup \@sanitize@url \@href}%
	\providecommand \@href[1]{\@@startlink{#1}\@@href}%
	\providecommand \@@href[1]{\endgroup#1\@@endlink}%
	\providecommand \@sanitize@url [0]{\catcode `\\12\catcode `\$12\catcode
		`\&12\catcode `\#12\catcode `\^12\catcode `\_12\catcode `\%12\relax}%
	\providecommand \@@startlink[1]{}%
	\providecommand \@@endlink[0]{}%
	\providecommand \url  [0]{\begingroup\@sanitize@url \@url }%
	\providecommand \@url [1]{\endgroup\@href {#1}{\urlprefix }}%
	\providecommand \urlprefix  [0]{URL }%
	\providecommand \Eprint [0]{\href }%
	\providecommand \doibase [0]{http://dx.doi.org/}%
	\providecommand \selectlanguage [0]{\@gobble}%
	\providecommand \bibinfo  [0]{\@secondoftwo}%
	\providecommand \bibfield  [0]{\@secondoftwo}%
	\providecommand \translation [1]{[#1]}%
	\providecommand \BibitemOpen [0]{}%
	\providecommand \bibitemStop [0]{}%
	\providecommand \bibitemNoStop [0]{.\EOS\space}%
	\providecommand \EOS [0]{\spacefactor3000\relax}%
	\providecommand \BibitemShut  [1]{\csname bibitem#1\endcsname}%
	\let\auto@bib@innerbib\@empty
	\bibitem [{\citenamefont {Hawking}(1976)}]{Hawking:1976ra}%
	\BibitemOpen
	\bibfield  {author} {\bibinfo {author} {\bibfnamefont {S.W.}\ \bibnamefont
			{Hawking}},\ }\bibfield  {title} {\enquote {\bibinfo {title} {{Breakdown of
					Predictability in Gravitational Collapse}},}\ }\href {\doibase
		10.1103/PhysRevD.14.2460} {\bibfield  {journal} {\bibinfo  {journal} {Phys.
				Rev. D}\ }\textbf {\bibinfo {volume} {14}},\ \bibinfo {pages} {2460--2473}
		(\bibinfo {year} {1976})}\BibitemShut {NoStop}%
	\bibitem [{\citenamefont {Hawking}(1975)}]{Hawking:1974sw}%
	\BibitemOpen
	\bibfield  {author} {\bibinfo {author} {\bibfnamefont {S.W.}\ \bibnamefont
			{Hawking}},\ }\bibfield  {title} {\enquote {\bibinfo {title} {{Particle
					Creation by Black Holes}},}\ }\href {\doibase 10.1007/BF02345020} {\bibfield
		{journal} {\bibinfo  {journal} {Commun. Math. Phys.}\ }\textbf {\bibinfo
			{volume} {43}},\ \bibinfo {pages} {199--220} (\bibinfo {year} {1975})},\
	\bibinfo {note} {[Erratum: Commun.Math.Phys. 46, 206 (1976)]}\BibitemShut
	{NoStop}%
	\bibitem [{\citenamefont {Grumiller}\ \emph {et~al.}(2002)\citenamefont
		{Grumiller}, \citenamefont {Kummer},\ and\ \citenamefont
		{Vassilevich}}]{Grumiller:2002nm}%
	\BibitemOpen
	\bibfield  {author} {\bibinfo {author} {\bibfnamefont {D.}~\bibnamefont
			{Grumiller}}, \bibinfo {author} {\bibfnamefont {W.}~\bibnamefont {Kummer}}, \
		and\ \bibinfo {author} {\bibfnamefont {D.V.}\ \bibnamefont {Vassilevich}},\
	}\bibfield  {title} {\enquote {\bibinfo {title} {{Dilaton gravity in
					two-dimensions}},}\ }\href {\doibase 10.1016/S0370-1573(02)00267-3}
	{\bibfield  {journal} {\bibinfo  {journal} {Phys. Rept.}\ }\textbf {\bibinfo
			{volume} {369}},\ \bibinfo {pages} {327--430} (\bibinfo {year} {2002})},\
	\Eprint {http://arxiv.org/abs/hep-th/0204253} {arXiv:hep-th/0204253}
	\BibitemShut {NoStop}%
	\bibitem [{\citenamefont {Jackiw}(1985)}]{Jackiw:1984je}%
	\BibitemOpen
	\bibfield  {author} {\bibinfo {author} {\bibfnamefont {R.}~\bibnamefont
			{Jackiw}},\ }\bibfield  {title} {\enquote {\bibinfo {title} {{Lower
					Dimensional Gravity}},}\ }\href {\doibase 10.1016/0550-3213(85)90448-1}
	{\bibfield  {journal} {\bibinfo  {journal} {Nucl. Phys. B}\ }\textbf
		{\bibinfo {volume} {252}},\ \bibinfo {pages} {343--356} (\bibinfo {year}
		{1985})}\BibitemShut {NoStop}%
	\bibitem [{\citenamefont {Teitelboim}(1983)}]{Teitelboim:1983ux}%
	\BibitemOpen
	\bibfield  {author} {\bibinfo {author} {\bibfnamefont {C.}~\bibnamefont
			{Teitelboim}},\ }\bibfield  {title} {\enquote {\bibinfo {title} {{Gravitation
					and Hamiltonian Structure in Two Space-Time Dimensions}},}\ }\href {\doibase
		10.1016/0370-2693(83)90012-6} {\bibfield  {journal} {\bibinfo  {journal}
			{Phys. Lett. B}\ }\textbf {\bibinfo {volume} {126}},\ \bibinfo {pages}
		{41--45} (\bibinfo {year} {1983})}\BibitemShut {NoStop}%
	\bibitem [{\citenamefont {Sachdev}\ and\ \citenamefont
		{Ye}(1993)}]{Sachdev:1992fk}%
	\BibitemOpen
	\bibfield  {author} {\bibinfo {author} {\bibfnamefont {Subir}\ \bibnamefont
			{Sachdev}}\ and\ \bibinfo {author} {\bibfnamefont {Jinwu}\ \bibnamefont
			{Ye}},\ }\bibfield  {title} {\enquote {\bibinfo {title} {{Gapless spin fluid
					ground state in a random, quantum Heisenberg magnet}},}\ }\href {\doibase
		10.1103/PhysRevLett.70.3339} {\bibfield  {journal} {\bibinfo  {journal}
			{Phys. Rev. Lett.}\ }\textbf {\bibinfo {volume} {70}},\ \bibinfo {pages}
		{3339} (\bibinfo {year} {1993})},\ \Eprint
	{http://arxiv.org/abs/cond-mat/9212030} {arXiv:cond-mat/9212030} \BibitemShut
	{NoStop}%
	\bibitem [{\citenamefont {Kitaev}()}]{kitaev:talk}%
	\BibitemOpen
	\bibfield  {author} {\bibinfo {author} {\bibfnamefont {Alexei}\ \bibnamefont
			{Kitaev}},\ }\href@noop {} {\enquote {\bibinfo {title} {A simple model of
				quantum holography},}\ }\bibinfo {note} {{Talk} given at the KITP Program:
		Entanglement in Strongly-Correlated Quantum Matter (April 6–July 2, 2015),
		University of California, Santa Barbara,
		{\href{https://online.kitp.ucsb.edu/online/entangled15/kitaev/}{Part 1}} and
		{\href{https://online.kitp.ucsb.edu/online/entangled15/kitaev2/}{Part 2}}
		available online.}\BibitemShut {Stop}%
	\bibitem [{\citenamefont {Maldacena}\ and\ \citenamefont
		{Stanford}(2016)}]{Maldacena:2016hyu}%
	\BibitemOpen
	\bibfield  {author} {\bibinfo {author} {\bibfnamefont {Juan}\ \bibnamefont
			{Maldacena}}\ and\ \bibinfo {author} {\bibfnamefont {Douglas}\ \bibnamefont
			{Stanford}},\ }\bibfield  {title} {\enquote {\bibinfo {title} {{Remarks on
					the Sachdev-Ye-Kitaev model}},}\ }\href {\doibase 10.1103/PhysRevD.94.106002}
	{\bibfield  {journal} {\bibinfo  {journal} {Phys. Rev. D}\ }\textbf {\bibinfo
			{volume} {94}},\ \bibinfo {pages} {106002} (\bibinfo {year} {2016})},\
	\Eprint {http://arxiv.org/abs/1604.07818} {arXiv:1604.07818 [hep-th]}
	\BibitemShut {NoStop}%
	\bibitem [{\citenamefont {Maldacena}(1999)}]{Maldacena:1997re}%
	\BibitemOpen
	\bibfield  {author} {\bibinfo {author} {\bibfnamefont {Juan~Martin}\
			\bibnamefont {Maldacena}},\ }\bibfield  {title} {\enquote {\bibinfo {title}
			{{The Large N limit of superconformal field theories and supergravity}},}\
	}\href {\doibase 10.1023/A:1026654312961} {\bibfield  {journal} {\bibinfo
			{journal} {Int. J. Theor. Phys.}\ }\textbf {\bibinfo {volume} {38}},\
		\bibinfo {pages} {1113--1133} (\bibinfo {year} {1999})},\ \Eprint
	{http://arxiv.org/abs/hep-th/9711200} {arXiv:hep-th/9711200} \BibitemShut
	{NoStop}%
	\bibitem [{\citenamefont {Almheiri}\ and\ \citenamefont
		{Polchinski}(2015)}]{Almheiri:2014cka}%
	\BibitemOpen
	\bibfield  {author} {\bibinfo {author} {\bibfnamefont {Ahmed}\ \bibnamefont
			{Almheiri}}\ and\ \bibinfo {author} {\bibfnamefont {Joseph}\ \bibnamefont
			{Polchinski}},\ }\bibfield  {title} {\enquote {\bibinfo {title} {{Models of
					AdS$_{2}$ backreaction and holography}},}\ }\href {\doibase
		10.1007/JHEP11(2015)014} {\bibfield  {journal} {\bibinfo  {journal} {JHEP}\
		}\textbf {\bibinfo {volume} {11}},\ \bibinfo {pages} {014} (\bibinfo {year}
		{2015})},\ \Eprint {http://arxiv.org/abs/1402.6334} {arXiv:1402.6334
		[hep-th]} \BibitemShut {NoStop}%
	\bibitem [{\citenamefont {Mann}\ \emph {et~al.}(1990)\citenamefont {Mann},
		\citenamefont {Shiekh},\ and\ \citenamefont {Tarasov}}]{Mann:1989gh}%
	\BibitemOpen
	\bibfield  {author} {\bibinfo {author} {\bibfnamefont {Robert~B.}\
			\bibnamefont {Mann}}, \bibinfo {author} {\bibfnamefont {A.}~\bibnamefont
			{Shiekh}}, \ and\ \bibinfo {author} {\bibfnamefont {L.}~\bibnamefont
			{Tarasov}},\ }\bibfield  {title} {\enquote {\bibinfo {title} {{Classical and
					Quantum Properties of Two-dimensional Black Holes}},}\ }\href {\doibase
		10.1016/0550-3213(90)90265-F} {\bibfield  {journal} {\bibinfo  {journal}
			{Nucl. Phys. B}\ }\textbf {\bibinfo {volume} {341}},\ \bibinfo {pages}
		{134--154} (\bibinfo {year} {1990})}\BibitemShut {NoStop}%
	\bibitem [{\citenamefont {Mandal}\ \emph {et~al.}(1991)\citenamefont {Mandal},
		\citenamefont {Sengupta},\ and\ \citenamefont {Wadia}}]{Mandal:1991tz}%
	\BibitemOpen
	\bibfield  {author} {\bibinfo {author} {\bibfnamefont {Gautam}\ \bibnamefont
			{Mandal}}, \bibinfo {author} {\bibfnamefont {Anirvan~M.}\ \bibnamefont
			{Sengupta}}, \ and\ \bibinfo {author} {\bibfnamefont {Spenta~R.}\
			\bibnamefont {Wadia}},\ }\bibfield  {title} {\enquote {\bibinfo {title}
			{{Classical solutions of two-dimensional string theory}},}\ }\href {\doibase
		10.1142/S0217732391001822} {\bibfield  {journal} {\bibinfo  {journal} {Mod.
				Phys. Lett. A}\ }\textbf {\bibinfo {volume} {6}},\ \bibinfo {pages}
		{1685--1692} (\bibinfo {year} {1991})}\BibitemShut {NoStop}%
	\bibitem [{\citenamefont {Navarro-Salas}\ and\ \citenamefont
		{Navarro}(2000)}]{NavarroSalas:1999up}%
	\BibitemOpen
	\bibfield  {author} {\bibinfo {author} {\bibfnamefont {J.}~\bibnamefont
			{Navarro-Salas}}\ and\ \bibinfo {author} {\bibfnamefont {P.}~\bibnamefont
			{Navarro}},\ }\bibfield  {title} {\enquote {\bibinfo {title}
			{{AdS$_2$/CFT$_1$ correspondence and near extremal black hole entropy}},}\
	}\href {\doibase 10.1016/S0550-3213(00)00165-6} {\bibfield  {journal}
		{\bibinfo  {journal} {Nucl. Phys. B}\ }\textbf {\bibinfo {volume} {579}},\
		\bibinfo {pages} {250--266} (\bibinfo {year} {2000})},\ \Eprint
	{http://arxiv.org/abs/hep-th/9910076} {arXiv:hep-th/9910076} \BibitemShut
	{NoStop}%
	\bibitem [{\citenamefont {Kettner}\ \emph {et~al.}(2004)\citenamefont
		{Kettner}, \citenamefont {Kunstatter},\ and\ \citenamefont
		{Medved}}]{Kettner:2004aw}%
	\BibitemOpen
	\bibfield  {author} {\bibinfo {author} {\bibfnamefont {Joanne}\ \bibnamefont
			{Kettner}}, \bibinfo {author} {\bibfnamefont {Gabor}\ \bibnamefont
			{Kunstatter}}, \ and\ \bibinfo {author} {\bibfnamefont {A.J.M.}\ \bibnamefont
			{Medved}},\ }\bibfield  {title} {\enquote {\bibinfo {title} {{Quasinormal
					modes for single horizon black holes in generic 2-d dilaton gravity}},}\
	}\href {\doibase 10.1088/0264-9381/21/23/002} {\bibfield  {journal} {\bibinfo
			{journal} {Class. Quant. Grav.}\ }\textbf {\bibinfo {volume} {21}},\
		\bibinfo {pages} {5317--5332} (\bibinfo {year} {2004})},\ \Eprint
	{http://arxiv.org/abs/gr-qc/0408042} {arXiv:gr-qc/0408042} \BibitemShut
	{NoStop}%
	\bibitem [{\citenamefont {Callan}\ \emph {et~al.}(1992)\citenamefont {Callan},
		\citenamefont {Giddings}, \citenamefont {Harvey},\ and\ \citenamefont
		{Strominger}}]{Callan:1992rs}%
	\BibitemOpen
	\bibfield  {author} {\bibinfo {author} {\bibfnamefont {Jr.}\ \bibnamefont
			{Callan}, \bibfnamefont {Curtis~G.}}, \bibinfo {author} {\bibfnamefont
			{Steven~B.}\ \bibnamefont {Giddings}}, \bibinfo {author} {\bibfnamefont
			{Jeffrey~A.}\ \bibnamefont {Harvey}}, \ and\ \bibinfo {author} {\bibfnamefont
			{Andrew}\ \bibnamefont {Strominger}},\ }\bibfield  {title} {\enquote
		{\bibinfo {title} {{Evanescent black holes}},}\ }\href {\doibase
		10.1103/PhysRevD.45.R1005} {\bibfield  {journal} {\bibinfo  {journal} {Phys.
				Rev. D}\ }\textbf {\bibinfo {volume} {45}},\ \bibinfo {pages} {1005}
		(\bibinfo {year} {1992})},\ \Eprint {http://arxiv.org/abs/hep-th/9111056}
	{arXiv:hep-th/9111056} \BibitemShut {NoStop}%
	\bibitem [{\citenamefont {Hollowood}\ and\ \citenamefont
		{Kumar}(2020)}]{Hollowood:2020cou}%
	\BibitemOpen
	\bibfield  {author} {\bibinfo {author} {\bibfnamefont {Timothy~J.}\
			\bibnamefont {Hollowood}}\ and\ \bibinfo {author} {\bibfnamefont {S.~Prem}\
			\bibnamefont {Kumar}},\ }\bibfield  {title} {\enquote {\bibinfo {title}
			{{Islands and Page Curves for Evaporating Black Holes in JT Gravity}},}\
	}\href {\doibase 10.1007/JHEP08(2020)094} {\bibfield  {journal} {\bibinfo
			{journal} {JHEP}\ }\textbf {\bibinfo {volume} {08}},\ \bibinfo {pages} {094}
		(\bibinfo {year} {2020})},\ \Eprint {http://arxiv.org/abs/2004.14944}
	{arXiv:2004.14944 [hep-th]} \BibitemShut {NoStop}%
	\bibitem [{\citenamefont {Almheiri}\ \emph {et~al.}(2020)\citenamefont
		{Almheiri}, \citenamefont {Hartman}, \citenamefont {Maldacena}, \citenamefont
		{Shaghoulian},\ and\ \citenamefont {Tajdini}}]{Almheiri:2019qdq}%
	\BibitemOpen
	\bibfield  {author} {\bibinfo {author} {\bibfnamefont {Ahmed}\ \bibnamefont
			{Almheiri}}, \bibinfo {author} {\bibfnamefont {Thomas}\ \bibnamefont
			{Hartman}}, \bibinfo {author} {\bibfnamefont {Juan}\ \bibnamefont
			{Maldacena}}, \bibinfo {author} {\bibfnamefont {Edgar}\ \bibnamefont
			{Shaghoulian}}, \ and\ \bibinfo {author} {\bibfnamefont {Amirhossein}\
			\bibnamefont {Tajdini}},\ }\bibfield  {title} {\enquote {\bibinfo {title}
			{{Replica Wormholes and the Entropy of Hawking Radiation}},}\ }\href
	{\doibase 10.1007/JHEP05(2020)013} {\bibfield  {journal} {\bibinfo  {journal}
			{JHEP}\ }\textbf {\bibinfo {volume} {05}},\ \bibinfo {pages} {013} (\bibinfo
		{year} {2020})},\ \Eprint {http://arxiv.org/abs/1911.12333} {arXiv:1911.12333
		[hep-th]} \BibitemShut {NoStop}%
	\bibitem [{\citenamefont {Moitra}\ \emph
		{et~al.}(2019{\natexlab{a}})\citenamefont {Moitra}, \citenamefont {Trivedi},\
		and\ \citenamefont {Vishal}}]{Moitra:2018jqs}%
	\BibitemOpen
	\bibfield  {author} {\bibinfo {author} {\bibfnamefont {Upamanyu}\
			\bibnamefont {Moitra}}, \bibinfo {author} {\bibfnamefont {Sandip~P.}\
			\bibnamefont {Trivedi}}, \ and\ \bibinfo {author} {\bibfnamefont
			{V.}~\bibnamefont {Vishal}},\ }\bibfield  {title} {\enquote {\bibinfo {title}
			{{Extremal and near-extremal black holes and near-CFT$_{1}$}},}\ }\href
	{\doibase 10.1007/JHEP07(2019)055} {\bibfield  {journal} {\bibinfo  {journal}
			{JHEP}\ }\textbf {\bibinfo {volume} {07}},\ \bibinfo {pages} {055} (\bibinfo
		{year} {2019}{\natexlab{a}})},\ \Eprint {http://arxiv.org/abs/1808.08239}
	{arXiv:1808.08239 [hep-th]} \BibitemShut {NoStop}%
	\bibitem [{\citenamefont {Nayak}\ \emph {et~al.}(2018)\citenamefont {Nayak},
		\citenamefont {Shukla}, \citenamefont {Soni}, \citenamefont {Trivedi},\ and\
		\citenamefont {Vishal}}]{Nayak:2018qej}%
	\BibitemOpen
	\bibfield  {author} {\bibinfo {author} {\bibfnamefont {Pranjal}\ \bibnamefont
			{Nayak}}, \bibinfo {author} {\bibfnamefont {Ashish}\ \bibnamefont {Shukla}},
		\bibinfo {author} {\bibfnamefont {Ronak~M.}\ \bibnamefont {Soni}}, \bibinfo
		{author} {\bibfnamefont {Sandip~P.}\ \bibnamefont {Trivedi}}, \ and\ \bibinfo
		{author} {\bibfnamefont {V.}~\bibnamefont {Vishal}},\ }\bibfield  {title}
	{\enquote {\bibinfo {title} {{On the Dynamics of Near-Extremal Black
					Holes}},}\ }\href {\doibase 10.1007/JHEP09(2018)048} {\bibfield  {journal}
		{\bibinfo  {journal} {JHEP}\ }\textbf {\bibinfo {volume} {09}},\ \bibinfo
		{pages} {048} (\bibinfo {year} {2018})},\ \Eprint
	{http://arxiv.org/abs/1802.09547} {arXiv:1802.09547 [hep-th]} \BibitemShut
	{NoStop}%
	\bibitem [{\citenamefont {Brown}\ \emph {et~al.}(2019)\citenamefont {Brown},
		\citenamefont {Gharibyan}, \citenamefont {Lin}, \citenamefont {Susskind},
		\citenamefont {Thorlacius},\ and\ \citenamefont {Zhao}}]{Brown:2018bms}%
	\BibitemOpen
	\bibfield  {author} {\bibinfo {author} {\bibfnamefont {Adam~R.}\ \bibnamefont
			{Brown}}, \bibinfo {author} {\bibfnamefont {Hrant}\ \bibnamefont
			{Gharibyan}}, \bibinfo {author} {\bibfnamefont {Henry~W.}\ \bibnamefont
			{Lin}}, \bibinfo {author} {\bibfnamefont {Leonard}\ \bibnamefont {Susskind}},
		\bibinfo {author} {\bibfnamefont {L\'arus}\ \bibnamefont {Thorlacius}}, \
		and\ \bibinfo {author} {\bibfnamefont {Ying}\ \bibnamefont {Zhao}},\
	}\bibfield  {title} {\enquote {\bibinfo {title} {{Complexity of
					Jackiw-Teitelboim gravity}},}\ }\href {\doibase 10.1103/PhysRevD.99.046016}
	{\bibfield  {journal} {\bibinfo  {journal} {Phys. Rev. D}\ }\textbf {\bibinfo
			{volume} {99}},\ \bibinfo {pages} {046016} (\bibinfo {year} {2019})},\
	\Eprint {http://arxiv.org/abs/1810.08741} {arXiv:1810.08741 [hep-th]}
	\BibitemShut {NoStop}%
	\bibitem [{\citenamefont {Moitra}\ \emph
		{et~al.}(2019{\natexlab{b}})\citenamefont {Moitra}, \citenamefont {Sake},
		\citenamefont {Trivedi},\ and\ \citenamefont {Vishal}}]{Moitra:2019bub}%
	\BibitemOpen
	\bibfield  {author} {\bibinfo {author} {\bibfnamefont {Upamanyu}\
			\bibnamefont {Moitra}}, \bibinfo {author} {\bibfnamefont {Sunil~Kumar}\
			\bibnamefont {Sake}}, \bibinfo {author} {\bibfnamefont {Sandip~P.}\
			\bibnamefont {Trivedi}}, \ and\ \bibinfo {author} {\bibfnamefont
			{V.}~\bibnamefont {Vishal}},\ }\bibfield  {title} {\enquote {\bibinfo {title}
			{{Jackiw-Teitelboim Gravity and Rotating Black Holes}},}\ }\href {\doibase
		10.1007/JHEP11(2019)047} {\bibfield  {journal} {\bibinfo  {journal} {JHEP}\
		}\textbf {\bibinfo {volume} {11}},\ \bibinfo {pages} {047} (\bibinfo {year}
		{2019}{\natexlab{b}})},\ \Eprint {http://arxiv.org/abs/1905.10378}
	{arXiv:1905.10378 [hep-th]} \BibitemShut {NoStop}%
	\bibitem [{\citenamefont {Kokkotas}\ and\ \citenamefont
		{Schmidt}(1999)}]{Kokkotas:1999bd}%
	\BibitemOpen
	\bibfield  {author} {\bibinfo {author} {\bibfnamefont {Kostas~D.}\
			\bibnamefont {Kokkotas}}\ and\ \bibinfo {author} {\bibfnamefont {Bernd~G.}\
			\bibnamefont {Schmidt}},\ }\bibfield  {title} {\enquote {\bibinfo {title}
			{{Quasinormal modes of stars and black holes}},}\ }\href {\doibase
		10.12942/lrr-1999-2} {\bibfield  {journal} {\bibinfo  {journal} {Living Rev.
				Rel.}\ }\textbf {\bibinfo {volume} {2}},\ \bibinfo {pages} {2} (\bibinfo
		{year} {1999})},\ \Eprint {http://arxiv.org/abs/gr-qc/9909058}
	{arXiv:gr-qc/9909058} \BibitemShut {NoStop}%
	\bibitem [{\citenamefont {Berti}\ \emph {et~al.}(2009)\citenamefont {Berti},
		\citenamefont {Cardoso},\ and\ \citenamefont {Starinets}}]{Berti:2009kk}%
	\BibitemOpen
	\bibfield  {author} {\bibinfo {author} {\bibfnamefont {Emanuele}\
			\bibnamefont {Berti}}, \bibinfo {author} {\bibfnamefont {Vitor}\ \bibnamefont
			{Cardoso}}, \ and\ \bibinfo {author} {\bibfnamefont {Andrei~O.}\ \bibnamefont
			{Starinets}},\ }\bibfield  {title} {\enquote {\bibinfo {title} {{Quasinormal
					modes of black holes and black branes}},}\ }\href {\doibase
		10.1088/0264-9381/26/16/163001} {\bibfield  {journal} {\bibinfo  {journal}
			{Class. Quant. Grav.}\ }\textbf {\bibinfo {volume} {26}},\ \bibinfo {pages}
		{163001} (\bibinfo {year} {2009})},\ \Eprint {http://arxiv.org/abs/0905.2975}
	{arXiv:0905.2975 [gr-qc]} \BibitemShut {NoStop}%
	\bibitem [{\citenamefont {Konoplya}\ and\ \citenamefont
		{Zhidenko}(2011)}]{Konoplya:2011qq}%
	\BibitemOpen
	\bibfield  {author} {\bibinfo {author} {\bibfnamefont {R.A.}\ \bibnamefont
			{Konoplya}}\ and\ \bibinfo {author} {\bibfnamefont {A.}~\bibnamefont
			{Zhidenko}},\ }\bibfield  {title} {\enquote {\bibinfo {title} {{Quasinormal
					modes of black holes: From astrophysics to string theory}},}\ }\href
	{\doibase 10.1103/RevModPhys.83.793} {\bibfield  {journal} {\bibinfo
			{journal} {Rev. Mod. Phys.}\ }\textbf {\bibinfo {volume} {83}},\ \bibinfo
		{pages} {793--836} (\bibinfo {year} {2011})},\ \Eprint
	{http://arxiv.org/abs/1102.4014} {arXiv:1102.4014 [gr-qc]} \BibitemShut
	{NoStop}%
	\bibitem [{\citenamefont {Horowitz}\ and\ \citenamefont
		{Hubeny}(2000)}]{Horowitz:1999jd}%
	\BibitemOpen
	\bibfield  {author} {\bibinfo {author} {\bibfnamefont {Gary~T.}\ \bibnamefont
			{Horowitz}}\ and\ \bibinfo {author} {\bibfnamefont {Veronika~E.}\
			\bibnamefont {Hubeny}},\ }\bibfield  {title} {\enquote {\bibinfo {title}
			{{Quasinormal modes of AdS black holes and the approach to thermal
					equilibrium}},}\ }\href {\doibase 10.1103/PhysRevD.62.024027} {\bibfield
		{journal} {\bibinfo  {journal} {Phys. Rev. D}\ }\textbf {\bibinfo {volume}
			{62}},\ \bibinfo {pages} {024027} (\bibinfo {year} {2000})},\ \Eprint
	{http://arxiv.org/abs/hep-th/9909056} {arXiv:hep-th/9909056} \BibitemShut
	{NoStop}%
	\bibitem [{\citenamefont {Damour}\ and\ \citenamefont
		{Ruffini}(1976)}]{Damour:1976jd}%
	\BibitemOpen
	\bibfield  {author} {\bibinfo {author} {\bibfnamefont {T.}~\bibnamefont
			{Damour}}\ and\ \bibinfo {author} {\bibfnamefont {R.}~\bibnamefont
			{Ruffini}},\ }\bibfield  {title} {\enquote {\bibinfo {title} {{Black Hole
					Evaporation in the Klein-Sauter-Heisenberg-Euler Formalism}},}\ }\href
	{\doibase 10.1103/PhysRevD.14.332} {\bibfield  {journal} {\bibinfo  {journal}
			{Phys. Rev. D}\ }\textbf {\bibinfo {volume} {14}},\ \bibinfo {pages}
		{332--334} (\bibinfo {year} {1976})}\BibitemShut {NoStop}%
	\bibitem [{\citenamefont {Sannan}(1988)}]{Sannan:1988eh}%
	\BibitemOpen
	\bibfield  {author} {\bibinfo {author} {\bibfnamefont {S.}~\bibnamefont
			{Sannan}},\ }\bibfield  {title} {\enquote {\bibinfo {title} {{Heuristic
					Derivation of the Probability Distributions of Particles Emitted by a Black
					Hole}},}\ }\href {\doibase 10.1007/BF00759183} {\bibfield  {journal}
		{\bibinfo  {journal} {Gen. Rel. Grav.}\ }\textbf {\bibinfo {volume} {20}},\
		\bibinfo {pages} {239--246} (\bibinfo {year} {1988})}\BibitemShut {NoStop}%
	\bibitem [{\citenamefont {Banados}\ \emph {et~al.}(1992)\citenamefont
		{Banados}, \citenamefont {Teitelboim},\ and\ \citenamefont
		{Zanelli}}]{Banados:1992wn}%
	\BibitemOpen
	\bibfield  {author} {\bibinfo {author} {\bibfnamefont {Maximo}\ \bibnamefont
			{Banados}}, \bibinfo {author} {\bibfnamefont {Claudio}\ \bibnamefont
			{Teitelboim}}, \ and\ \bibinfo {author} {\bibfnamefont {Jorge}\ \bibnamefont
			{Zanelli}},\ }\bibfield  {title} {\enquote {\bibinfo {title} {{The Black hole
					in three-dimensional space-time}},}\ }\href {\doibase
		10.1103/PhysRevLett.69.1849} {\bibfield  {journal} {\bibinfo  {journal}
			{Phys. Rev. Lett.}\ }\textbf {\bibinfo {volume} {69}},\ \bibinfo {pages}
		{1849--1851} (\bibinfo {year} {1992})},\ \Eprint
	{http://arxiv.org/abs/hep-th/9204099} {arXiv:hep-th/9204099} \BibitemShut
	{NoStop}%
	\bibitem [{\citenamefont {Dias}\ \emph {et~al.}(2019)\citenamefont {Dias},
		\citenamefont {Reall},\ and\ \citenamefont {Santos}}]{Dias:2019ery}%
	\BibitemOpen
	\bibfield  {author} {\bibinfo {author} {\bibfnamefont {Oscar~J.C.}\
			\bibnamefont {Dias}}, \bibinfo {author} {\bibfnamefont {Harvey~S.}\
			\bibnamefont {Reall}}, \ and\ \bibinfo {author} {\bibfnamefont {Jorge~E.}\
			\bibnamefont {Santos}},\ }\bibfield  {title} {\enquote {\bibinfo {title}
			{{The BTZ black hole violates strong cosmic censorship}},}\ }\href {\doibase
		10.1007/JHEP12(2019)097} {\bibfield  {journal} {\bibinfo  {journal} {JHEP}\
		}\textbf {\bibinfo {volume} {12}},\ \bibinfo {pages} {097} (\bibinfo {year}
		{2019})},\ \Eprint {http://arxiv.org/abs/1906.08265} {arXiv:1906.08265
		[hep-th]} \BibitemShut {NoStop}%
	\bibitem [{\citenamefont {Bhattacharjee}\ \emph {et~al.}(2020)\citenamefont
		{Bhattacharjee}, \citenamefont {Kumar},\ and\ \citenamefont
		{Sarkar}}]{Bhattacharjee:2020gbo}%
	\BibitemOpen
	\bibfield  {author} {\bibinfo {author} {\bibfnamefont {Srijit}\ \bibnamefont
			{Bhattacharjee}}, \bibinfo {author} {\bibfnamefont {Shailesh}\ \bibnamefont
			{Kumar}}, \ and\ \bibinfo {author} {\bibfnamefont {Subhodeep}\ \bibnamefont
			{Sarkar}},\ }\bibfield  {title} {\enquote {\bibinfo {title} {{Mass inflation
					and strong cosmic censorship in a nonextreme BTZ black hole}},}\ }\href
	{\doibase 10.1103/PhysRevD.102.044030} {\bibfield  {journal} {\bibinfo
			{journal} {Phys. Rev. D}\ }\textbf {\bibinfo {volume} {102}},\ \bibinfo
		{pages} {044030} (\bibinfo {year} {2020})},\ \Eprint
	{http://arxiv.org/abs/2005.09705} {arXiv:2005.09705 [gr-qc]} \BibitemShut
	{NoStop}%
	\bibitem [{\citenamefont {Moitra}(2020)}]{Moitra:2020ojo}%
	\BibitemOpen
	\bibfield  {author} {\bibinfo {author} {\bibfnamefont {Upamanyu}\
			\bibnamefont {Moitra}},\ }\bibfield  {title} {\enquote {\bibinfo {title} {{On
					Strong Cosmic Censorship in Two Dimensions}},}\ }\href@noop {} {\  (\bibinfo
		{year} {2020})},\ \Eprint {http://arxiv.org/abs/2011.03499} {arXiv:2011.03499
		[hep-th]} \BibitemShut {NoStop}%
	\bibitem [{\citenamefont {Louis-Martinez}\ \emph {et~al.}(1994)\citenamefont
		{Louis-Martinez}, \citenamefont {Gegenberg},\ and\ \citenamefont
		{Kunstatter}}]{LouisMartinez:1993eh}%
	\BibitemOpen
	\bibfield  {author} {\bibinfo {author} {\bibfnamefont {D.}~\bibnamefont
			{Louis-Martinez}}, \bibinfo {author} {\bibfnamefont {J.}~\bibnamefont
			{Gegenberg}}, \ and\ \bibinfo {author} {\bibfnamefont {G.}~\bibnamefont
			{Kunstatter}},\ }\bibfield  {title} {\enquote {\bibinfo {title} {{Exact Dirac
					quantization of all 2-D dilaton gravity theories}},}\ }\href {\doibase
		10.1016/0370-2693(94)90463-4} {\bibfield  {journal} {\bibinfo  {journal}
			{Phys. Lett. B}\ }\textbf {\bibinfo {volume} {321}},\ \bibinfo {pages}
		{193--198} (\bibinfo {year} {1994})},\ \Eprint
	{http://arxiv.org/abs/gr-qc/9309018} {arXiv:gr-qc/9309018} \BibitemShut
	{NoStop}%
	\bibitem [{\citenamefont {Louis-Martinez}\ and\ \citenamefont
		{Kunstatter}(1994)}]{LouisMartinez:1993cc}%
	\BibitemOpen
	\bibfield  {author} {\bibinfo {author} {\bibfnamefont {D.}~\bibnamefont
			{Louis-Martinez}}\ and\ \bibinfo {author} {\bibfnamefont {G.}~\bibnamefont
			{Kunstatter}},\ }\bibfield  {title} {\enquote {\bibinfo {title} {{On
					Birckhoff's theorem in 2-D dilaton gravity}},}\ }\href {\doibase
		10.1103/PhysRevD.49.5227} {\bibfield  {journal} {\bibinfo  {journal} {Phys.
				Rev. D}\ }\textbf {\bibinfo {volume} {49}},\ \bibinfo {pages} {5227--5230}
		(\bibinfo {year} {1994})}\BibitemShut {NoStop}%
	\bibitem [{\citenamefont {Louis-Martinez}\ and\ \citenamefont
		{Kunstatter}(1995)}]{LouisMartinez:1995rq}%
	\BibitemOpen
	\bibfield  {author} {\bibinfo {author} {\bibfnamefont {D.}~\bibnamefont
			{Louis-Martinez}}\ and\ \bibinfo {author} {\bibfnamefont {G.}~\bibnamefont
			{Kunstatter}},\ }\bibfield  {title} {\enquote {\bibinfo {title}
			{{Two-dimensional dilaton gravity coupled to an Abelian gauged field}},}\
	}\href {\doibase 10.1103/PhysRevD.52.3494} {\bibfield  {journal} {\bibinfo
			{journal} {Phys. Rev. D}\ }\textbf {\bibinfo {volume} {52}},\ \bibinfo
		{pages} {3494--3505} (\bibinfo {year} {1995})},\ \Eprint
	{http://arxiv.org/abs/gr-qc/9503016} {arXiv:gr-qc/9503016} \BibitemShut
	{NoStop}%
	\bibitem [{\citenamefont {Achucarro}\ and\ \citenamefont
		{Ortiz}(1993)}]{Achucarro:1993fd}%
	\BibitemOpen
	\bibfield  {author} {\bibinfo {author} {\bibfnamefont {Ana}\ \bibnamefont
			{Achucarro}}\ and\ \bibinfo {author} {\bibfnamefont {Miguel~E.}\ \bibnamefont
			{Ortiz}},\ }\bibfield  {title} {\enquote {\bibinfo {title} {{Relating black
					holes in two-dimensions and three-dimensions}},}\ }\href {\doibase
		10.1103/PhysRevD.48.3600} {\bibfield  {journal} {\bibinfo  {journal} {Phys.
				Rev. D}\ }\textbf {\bibinfo {volume} {48}},\ \bibinfo {pages} {3600--3605}
		(\bibinfo {year} {1993})},\ \Eprint {http://arxiv.org/abs/hep-th/9304068}
	{arXiv:hep-th/9304068} \BibitemShut {NoStop}%
	\bibitem [{\citenamefont {Gegenberg}\ \emph {et~al.}(1995)\citenamefont
		{Gegenberg}, \citenamefont {Kunstatter},\ and\ \citenamefont
		{Louis-Martinez}}]{Gegenberg:1994pv}%
	\BibitemOpen
	\bibfield  {author} {\bibinfo {author} {\bibfnamefont {J.}~\bibnamefont
			{Gegenberg}}, \bibinfo {author} {\bibfnamefont {G.}~\bibnamefont
			{Kunstatter}}, \ and\ \bibinfo {author} {\bibfnamefont {D.}~\bibnamefont
			{Louis-Martinez}},\ }\bibfield  {title} {\enquote {\bibinfo {title}
			{{Observables for two-dimensional black holes}},}\ }\href {\doibase
		10.1103/PhysRevD.51.1781} {\bibfield  {journal} {\bibinfo  {journal} {Phys.
				Rev. D}\ }\textbf {\bibinfo {volume} {51}},\ \bibinfo {pages} {1781--1786}
		(\bibinfo {year} {1995})},\ \Eprint {http://arxiv.org/abs/gr-qc/9408015}
	{arXiv:gr-qc/9408015} \BibitemShut {NoStop}%
	\bibitem [{\citenamefont {Kunstatter}\ \emph {et~al.}(1998)\citenamefont
		{Kunstatter}, \citenamefont {Petryk},\ and\ \citenamefont
		{Shelemy}}]{Kunstatter:1997my}%
	\BibitemOpen
	\bibfield  {author} {\bibinfo {author} {\bibfnamefont {G.}~\bibnamefont
			{Kunstatter}}, \bibinfo {author} {\bibfnamefont {R.}~\bibnamefont {Petryk}},
		\ and\ \bibinfo {author} {\bibfnamefont {S.}~\bibnamefont {Shelemy}},\
	}\bibfield  {title} {\enquote {\bibinfo {title} {{Hamiltonian thermodynamics
					of black holes in generic 2-D dilaton gravity}},}\ }\href {\doibase
		10.1103/PhysRevD.57.3537} {\bibfield  {journal} {\bibinfo  {journal} {Phys.
				Rev. D}\ }\textbf {\bibinfo {volume} {57}},\ \bibinfo {pages} {3537--3547}
		(\bibinfo {year} {1998})},\ \Eprint {http://arxiv.org/abs/gr-qc/9709043}
	{arXiv:gr-qc/9709043} \BibitemShut {NoStop}%
	\bibitem [{\citenamefont {Cordero}\ \emph {et~al.}(2012)\citenamefont
		{Cordero}, \citenamefont {Lopez-Ortega},\ and\ \citenamefont
		{Vega-Acevedo}}]{Cordero:2012je}%
	\BibitemOpen
	\bibfield  {author} {\bibinfo {author} {\bibfnamefont {R.}~\bibnamefont
			{Cordero}}, \bibinfo {author} {\bibfnamefont {A.}~\bibnamefont
			{Lopez-Ortega}}, \ and\ \bibinfo {author} {\bibfnamefont {I.}~\bibnamefont
			{Vega-Acevedo}},\ }\bibfield  {title} {\enquote {\bibinfo {title}
			{{Quasinormal frequencies of asymptotically anti-de Sitter black holes in two
					dimensions}},}\ }\href {\doibase 10.1007/s10714-011-1316-1} {\bibfield
		{journal} {\bibinfo  {journal} {Gen. Rel. Grav.}\ }\textbf {\bibinfo {volume}
			{44}},\ \bibinfo {pages} {917--940} (\bibinfo {year} {2012})},\ \Eprint
	{http://arxiv.org/abs/1201.3605} {arXiv:1201.3605 [gr-qc]} \BibitemShut
	{NoStop}%
	\bibitem [{\citenamefont {Cardoso}\ and\ \citenamefont
		{Lemos}(2001{\natexlab{a}})}]{Cardoso:2001hn}%
	\BibitemOpen
	\bibfield  {author} {\bibinfo {author} {\bibfnamefont {Vitor}\ \bibnamefont
			{Cardoso}}\ and\ \bibinfo {author} {\bibfnamefont {Jose~P.S.}\ \bibnamefont
			{Lemos}},\ }\bibfield  {title} {\enquote {\bibinfo {title} {{Scalar,
					electromagnetic and Weyl perturbations of BTZ black holes: Quasinormal
					modes}},}\ }\href {\doibase 10.1103/PhysRevD.63.124015} {\bibfield  {journal}
		{\bibinfo  {journal} {Phys. Rev. D}\ }\textbf {\bibinfo {volume} {63}},\
		\bibinfo {pages} {124015} (\bibinfo {year} {2001}{\natexlab{a}})},\ \Eprint
	{http://arxiv.org/abs/gr-qc/0101052} {arXiv:gr-qc/0101052} \BibitemShut
	{NoStop}%
	\bibitem [{\citenamefont {Du}\ \emph {et~al.}(2004)\citenamefont {Du},
		\citenamefont {Wang},\ and\ \citenamefont {Su}}]{Du:2004jt}%
	\BibitemOpen
	\bibfield  {author} {\bibinfo {author} {\bibfnamefont {Da-Ping}\ \bibnamefont
			{Du}}, \bibinfo {author} {\bibfnamefont {Bin}\ \bibnamefont {Wang}}, \ and\
		\bibinfo {author} {\bibfnamefont {Ru-Keng}\ \bibnamefont {Su}},\ }\bibfield
	{title} {\enquote {\bibinfo {title} {{Quasinormal modes in pure de Sitter
					space-times}},}\ }\href {\doibase 10.1103/PhysRevD.70.064024} {\bibfield
		{journal} {\bibinfo  {journal} {Phys. Rev. D}\ }\textbf {\bibinfo {volume}
			{70}},\ \bibinfo {pages} {064024} (\bibinfo {year} {2004})},\ \Eprint
	{http://arxiv.org/abs/hep-th/0404047} {arXiv:hep-th/0404047} \BibitemShut
	{NoStop}%
	\bibitem [{\citenamefont {Fabris}\ \emph {et~al.}(2020)\citenamefont {Fabris},
		\citenamefont {Richarte},\ and\ \citenamefont {Saa}}]{Fabris:2020kog}%
	\BibitemOpen
	\bibfield  {author} {\bibinfo {author} {\bibfnamefont {J\'ulio~C.}\
			\bibnamefont {Fabris}}, \bibinfo {author} {\bibfnamefont {Mart\'\i{}n~G.}\
			\bibnamefont {Richarte}}, \ and\ \bibinfo {author} {\bibfnamefont {Alberto}\
			\bibnamefont {Saa}},\ }\bibfield  {title} {\enquote {\bibinfo {title}
			{{Quasinormal modes and self-adjoint extensions of the Schroedinger
					operator}},}\ }\href@noop {} {\  (\bibinfo {year} {2020})},\ \Eprint
	{http://arxiv.org/abs/2010.10674} {arXiv:2010.10674 [gr-qc]} \BibitemShut
	{NoStop}%
	\bibitem [{\citenamefont {Nikiforov}\ and\ \citenamefont
		{Uvarov}(1988)}]{Nikiforov1988}%
	\BibitemOpen
	\bibfield  {author} {\bibinfo {author} {\bibfnamefont {Arnold~F.}\
			\bibnamefont {Nikiforov}}\ and\ \bibinfo {author} {\bibfnamefont
			{Vasilii~B.}\ \bibnamefont {Uvarov}},\ }\enquote {\bibinfo {title}
		{Hypergeometric functions},}\ in\ \href {\doibase
		10.1007/978-1-4757-1595-8_4} {\emph {\bibinfo {booktitle} {Special Functions
				of Mathematical Physics: A Unified Introduction with Applications}}}\
	(\bibinfo  {publisher} {Birkh{\"a}user Boston},\ \bibinfo {address} {Boston,
		MA},\ \bibinfo {year} {1988})\ pp.\ \bibinfo {pages} {253--294}\BibitemShut
	{NoStop}%
	\bibitem [{\citenamefont {Birmingham}(2001)}]{Birmingham:2001hc}%
	\BibitemOpen
	\bibfield  {author} {\bibinfo {author} {\bibfnamefont {Danny}\ \bibnamefont
			{Birmingham}},\ }\bibfield  {title} {\enquote {\bibinfo {title} {{Choptuik
					scaling and quasinormal modes in the AdS / CFT correspondence}},}\ }\href
	{\doibase 10.1103/PhysRevD.64.064024} {\bibfield  {journal} {\bibinfo
			{journal} {Phys. Rev. D}\ }\textbf {\bibinfo {volume} {64}},\ \bibinfo
		{pages} {064024} (\bibinfo {year} {2001})},\ \Eprint
	{http://arxiv.org/abs/hep-th/0101194} {arXiv:hep-th/0101194} \BibitemShut
	{NoStop}%
	\bibitem [{\citenamefont {Abramowitz}\ and\ \citenamefont
		{Stegun}(1964)}]{abramowitz+stegun}%
	\BibitemOpen
	\bibfield  {author} {\bibinfo {author} {\bibfnamefont {Milton}\ \bibnamefont
			{Abramowitz}}\ and\ \bibinfo {author} {\bibfnamefont {Irene~A.}\ \bibnamefont
			{Stegun}},\ }\href@noop {} {\emph {\bibinfo {title} {Handbook of Mathematical
				Functions with Formulas, Graphs, and Mathematical Tables}}},\ \bibinfo
	{edition} {ninth dover printing, tenth gpo printing}\ ed.\ (\bibinfo
	{publisher} {Dover},\ \bibinfo {address} {New York},\ \bibinfo {year}
	{1964})\BibitemShut {NoStop}%
	\bibitem [{\citenamefont {Cardoso}\ and\ \citenamefont
		{Lemos}(2001{\natexlab{b}})}]{Cardoso:2001bb}%
	\BibitemOpen
	\bibfield  {author} {\bibinfo {author} {\bibfnamefont {Vitor}\ \bibnamefont
			{Cardoso}}\ and\ \bibinfo {author} {\bibfnamefont {Jose~P.S.}\ \bibnamefont
			{Lemos}},\ }\bibfield  {title} {\enquote {\bibinfo {title} {{Quasinormal
					modes of Schwarzschild anti-de Sitter black holes: Electromagnetic and
					gravitational perturbations}},}\ }\href {\doibase 10.1103/PhysRevD.64.084017}
	{\bibfield  {journal} {\bibinfo  {journal} {Phys. Rev. D}\ }\textbf {\bibinfo
			{volume} {64}},\ \bibinfo {pages} {084017} (\bibinfo {year}
		{2001}{\natexlab{b}})},\ \Eprint {http://arxiv.org/abs/gr-qc/0105103}
	{arXiv:gr-qc/0105103} \BibitemShut {NoStop}%
	\bibitem [{\citenamefont {Cardoso}\ \emph {et~al.}(2003)\citenamefont
		{Cardoso}, \citenamefont {Konoplya},\ and\ \citenamefont
		{Lemos}}]{Cardoso:2003cj}%
	\BibitemOpen
	\bibfield  {author} {\bibinfo {author} {\bibfnamefont {Vitor}\ \bibnamefont
			{Cardoso}}, \bibinfo {author} {\bibfnamefont {Roman}\ \bibnamefont
			{Konoplya}}, \ and\ \bibinfo {author} {\bibfnamefont {Jose~P.S.}\
			\bibnamefont {Lemos}},\ }\bibfield  {title} {\enquote {\bibinfo {title}
			{{Quasinormal frequencies of Schwarzschild black holes in anti-de Sitter
					space-times: A Complete study on the asymptotic behavior}},}\ }\href
	{\doibase 10.1103/PhysRevD.68.044024} {\bibfield  {journal} {\bibinfo
			{journal} {Phys. Rev. D}\ }\textbf {\bibinfo {volume} {68}},\ \bibinfo
		{pages} {044024} (\bibinfo {year} {2003})},\ \Eprint
	{http://arxiv.org/abs/gr-qc/0305037} {arXiv:gr-qc/0305037} \BibitemShut
	{NoStop}%
	\bibitem [{\citenamefont {Hod}(1998)}]{Hod:1998vk}%
	\BibitemOpen
	\bibfield  {author} {\bibinfo {author} {\bibfnamefont {Shahar}\ \bibnamefont
			{Hod}},\ }\bibfield  {title} {\enquote {\bibinfo {title} {{Bohr's
					correspondence principle and the area spectrum of quantum black holes}},}\
	}\href {\doibase 10.1103/PhysRevLett.81.4293} {\bibfield  {journal} {\bibinfo
			{journal} {Phys. Rev. Lett.}\ }\textbf {\bibinfo {volume} {81}},\ \bibinfo
		{pages} {4293} (\bibinfo {year} {1998})},\ \Eprint
	{http://arxiv.org/abs/gr-qc/9812002} {arXiv:gr-qc/9812002} \BibitemShut
	{NoStop}%
	\bibitem [{\citenamefont {Fiziev}(2009)}]{Fiziev_2009}%
	\BibitemOpen
	\bibfield  {author} {\bibinfo {author} {\bibfnamefont {Plamen~P}\
			\bibnamefont {Fiziev}},\ }\bibfield  {title} {\enquote {\bibinfo {title}
			{Novel relations and new properties of confluent heun’s functions and their
				derivatives of arbitrary order},}\ }\href {\doibase
		10.1088/1751-8113/43/3/035203} {\bibfield  {journal} {\bibinfo  {journal}
			{Journal of Physics A: Mathematical and Theoretical}\ }\textbf {\bibinfo
			{volume} {43}},\ \bibinfo {pages} {035203} (\bibinfo {year}
		{2009})}\BibitemShut {NoStop}%
	\bibitem [{\citenamefont {Bhattacharyya}\ \emph {et~al.}(2020)\citenamefont
		{Bhattacharyya}, \citenamefont {Hilditch}, \citenamefont {Rajesh~Nayak},
		\citenamefont {R\"uter},\ and\ \citenamefont
		{Br\"ugmann}}]{Bhattacharyya:2020zkc}%
	\BibitemOpen
	\bibfield  {author} {\bibinfo {author} {\bibfnamefont {Maitraya~K.}\
			\bibnamefont {Bhattacharyya}}, \bibinfo {author} {\bibfnamefont {David}\
			\bibnamefont {Hilditch}}, \bibinfo {author} {\bibfnamefont {K.}~\bibnamefont
			{Rajesh~Nayak}}, \bibinfo {author} {\bibfnamefont {Hannes~R.}\ \bibnamefont
			{R\"uter}}, \ and\ \bibinfo {author} {\bibfnamefont {Bernd}\ \bibnamefont
			{Br\"ugmann}},\ }\bibfield  {title} {\enquote {\bibinfo {title} {{Analytical
					and numerical treatment of perturbed black holes in horizon-penetrating
					coordinates}},}\ }\href {\doibase 10.1103/PhysRevD.102.024039} {\bibfield
		{journal} {\bibinfo  {journal} {Phys. Rev. D}\ }\textbf {\bibinfo {volume}
			{102}},\ \bibinfo {pages} {024039} (\bibinfo {year} {2020})},\ \Eprint
	{http://arxiv.org/abs/2004.02558} {arXiv:2004.02558 [gr-qc]} \BibitemShut
	{NoStop}%
	\bibitem [{\citenamefont {Vieira}\ \emph {et~al.}(2015)\citenamefont {Vieira},
		\citenamefont {Bezerra},\ and\ \citenamefont {Costa}}]{Vieira:2014iea}%
	\BibitemOpen
	\bibfield  {author} {\bibinfo {author} {\bibfnamefont {H.S.}\ \bibnamefont
			{Vieira}}, \bibinfo {author} {\bibfnamefont {V.B.}\ \bibnamefont {Bezerra}},
		\ and\ \bibinfo {author} {\bibfnamefont {Andr\'e~A.}\ \bibnamefont {Costa}},\
	}\bibfield  {title} {\enquote {\bibinfo {title} {{Scalar fields in the
					Lense-Thirring background with a cosmic string and Hawking radiation}},}\
	}\href {\doibase 10.1209/0295-5075/109/60006} {\bibfield  {journal} {\bibinfo
			{journal} {EPL}\ }\textbf {\bibinfo {volume} {109}},\ \bibinfo {pages}
		{60006} (\bibinfo {year} {2015})},\ \Eprint {http://arxiv.org/abs/1405.7846}
	{arXiv:1405.7846 [gr-qc]} \BibitemShut {NoStop}%
	\bibitem [{\citenamefont {Sakalli}(2016)}]{Sakalli:2016fif}%
	\BibitemOpen
	\bibfield  {author} {\bibinfo {author} {\bibfnamefont {I.}~\bibnamefont
			{Sakalli}},\ }\bibfield  {title} {\enquote {\bibinfo {title} {{Analytical
					solutions in rotating linear dilaton black holes: resonant frequencies,
					quantization, greybody factor, and Hawking radiation}},}\ }\href {\doibase
		10.1103/PhysRevD.94.084040} {\bibfield  {journal} {\bibinfo  {journal} {Phys.
				Rev. D}\ }\textbf {\bibinfo {volume} {94}},\ \bibinfo {pages} {084040}
		(\bibinfo {year} {2016})},\ \Eprint {http://arxiv.org/abs/1606.00896}
	{arXiv:1606.00896 [gr-qc]} \BibitemShut {NoStop}%
	\bibitem [{\citenamefont {Breitenlohner}\ and\ \citenamefont
		{Freedman}(1982)}]{Breitenlohner:1982jf}%
	\BibitemOpen
	\bibfield  {author} {\bibinfo {author} {\bibfnamefont {Peter}\ \bibnamefont
			{Breitenlohner}}\ and\ \bibinfo {author} {\bibfnamefont {Daniel~Z.}\
			\bibnamefont {Freedman}},\ }\bibfield  {title} {\enquote {\bibinfo {title}
			{{Stability in Gauged Extended Supergravity}},}\ }\href {\doibase
		10.1016/0003-4916(82)90116-6} {\bibfield  {journal} {\bibinfo  {journal}
			{Annals Phys.}\ }\textbf {\bibinfo {volume} {144}},\ \bibinfo {pages} {249}
		(\bibinfo {year} {1982})}\BibitemShut {NoStop}%
	\bibitem [{\citenamefont {Dias}\ \emph {et~al.}(2018)\citenamefont {Dias},
		\citenamefont {Reall},\ and\ \citenamefont {Santos}}]{Dias:2018etb}%
	\BibitemOpen
	\bibfield  {author} {\bibinfo {author} {\bibfnamefont {Oscar~J.C.}\
			\bibnamefont {Dias}}, \bibinfo {author} {\bibfnamefont {Harvey~S.}\
			\bibnamefont {Reall}}, \ and\ \bibinfo {author} {\bibfnamefont {Jorge~E.}\
			\bibnamefont {Santos}},\ }\bibfield  {title} {\enquote {\bibinfo {title}
			{{Strong cosmic censorship: taking the rough with the smooth}},}\ }\href
	{\doibase 10.1007/JHEP10(2018)001} {\bibfield  {journal} {\bibinfo  {journal}
			{JHEP}\ }\textbf {\bibinfo {volume} {10}},\ \bibinfo {pages} {001} (\bibinfo
		{year} {2018})},\ \Eprint {http://arxiv.org/abs/1808.02895} {arXiv:1808.02895
		[gr-qc]} \BibitemShut {NoStop}%
	\bibitem [{\citenamefont {Marolf}\ and\ \citenamefont
		{Ori}(2012)}]{Marolf:2011dj}%
	\BibitemOpen
	\bibfield  {author} {\bibinfo {author} {\bibfnamefont {Donald}\ \bibnamefont
			{Marolf}}\ and\ \bibinfo {author} {\bibfnamefont {Amos}\ \bibnamefont
			{Ori}},\ }\bibfield  {title} {\enquote {\bibinfo {title} {{Outgoing
					gravitational shock-wave at the inner horizon: The late-time limit of black
					hole interiors}},}\ }\href {\doibase 10.1103/PhysRevD.86.124026} {\bibfield
		{journal} {\bibinfo  {journal} {Phys. Rev. D}\ }\textbf {\bibinfo {volume}
			{86}},\ \bibinfo {pages} {124026} (\bibinfo {year} {2012})},\ \Eprint
	{http://arxiv.org/abs/1109.5139} {arXiv:1109.5139 [gr-qc]} \BibitemShut
	{NoStop}%
	\bibitem [{\citenamefont {Balasubramanian}\ and\ \citenamefont
		{Levi}(2004)}]{Balasubramanian:2004zu}%
	\BibitemOpen
	\bibfield  {author} {\bibinfo {author} {\bibfnamefont {Vijay}\ \bibnamefont
			{Balasubramanian}}\ and\ \bibinfo {author} {\bibfnamefont {Thomas~S.}\
			\bibnamefont {Levi}},\ }\bibfield  {title} {\enquote {\bibinfo {title}
			{{Beyond the veil: Inner horizon instability and holography}},}\ }\href
	{\doibase 10.1103/PhysRevD.70.106005} {\bibfield  {journal} {\bibinfo
			{journal} {Phys. Rev. D}\ }\textbf {\bibinfo {volume} {70}},\ \bibinfo
		{pages} {106005} (\bibinfo {year} {2004})},\ \Eprint
	{http://arxiv.org/abs/hep-th/0405048} {arXiv:hep-th/0405048} \BibitemShut
	{NoStop}%
	\bibitem [{\citenamefont {Pandya}\ and\ \citenamefont
		{Pretorius}(2020)}]{Pandya:2020ejc}%
	\BibitemOpen
	\bibfield  {author} {\bibinfo {author} {\bibfnamefont {Alex}\ \bibnamefont
			{Pandya}}\ and\ \bibinfo {author} {\bibfnamefont {Frans}\ \bibnamefont
			{Pretorius}},\ }\bibfield  {title} {\enquote {\bibinfo {title} {{The rotating
					black hole interior: Insights from gravitational collapse in $AdS_3$
					spacetime}},}\ }\href {\doibase 10.1103/PhysRevD.101.104026} {\bibfield
		{journal} {\bibinfo  {journal} {Phys. Rev. D}\ }\textbf {\bibinfo {volume}
			{101}},\ \bibinfo {pages} {104026} (\bibinfo {year} {2020})},\ \Eprint
	{http://arxiv.org/abs/2002.07130} {arXiv:2002.07130 [gr-qc]} \BibitemShut
	{NoStop}%
	\bibitem [{\citenamefont {Birrell}\ and\ \citenamefont
		{Davies}(1978)}]{Birrell:1978th}%
	\BibitemOpen
	\bibfield  {author} {\bibinfo {author} {\bibfnamefont {N.D.}\ \bibnamefont
			{Birrell}}\ and\ \bibinfo {author} {\bibfnamefont {P.C.W.}\ \bibnamefont
			{Davies}},\ }\bibfield  {title} {\enquote {\bibinfo {title} {{On falling
					through a black hole into another universe}},}\ }\href {\doibase
		10.1038/272035a0} {\bibfield  {journal} {\bibinfo  {journal} {Nature}\
		}\textbf {\bibinfo {volume} {272}},\ \bibinfo {pages} {35} (\bibinfo {year}
		{1978})}\BibitemShut {NoStop}%
	\bibitem [{\citenamefont {Hollands}\ \emph {et~al.}(2020)\citenamefont
		{Hollands}, \citenamefont {Wald},\ and\ \citenamefont
		{Zahn}}]{Hollands:2019whz}%
	\BibitemOpen
	\bibfield  {author} {\bibinfo {author} {\bibfnamefont {Stefan}\ \bibnamefont
			{Hollands}}, \bibinfo {author} {\bibfnamefont {Robert~M.}\ \bibnamefont
			{Wald}}, \ and\ \bibinfo {author} {\bibfnamefont {Jochen}\ \bibnamefont
			{Zahn}},\ }\bibfield  {title} {\enquote {\bibinfo {title} {{Quantum
					instability of the Cauchy horizon in
					Reissner\textendash{}Nordstr\"om\textendash{}deSitter spacetime}},}\ }\href
	{\doibase 10.1088/1361-6382/ab8052} {\bibfield  {journal} {\bibinfo
			{journal} {Class. Quant. Grav.}\ }\textbf {\bibinfo {volume} {37}},\ \bibinfo
		{pages} {115009} (\bibinfo {year} {2020})},\ \Eprint
	{http://arxiv.org/abs/1912.06047} {arXiv:1912.06047 [gr-qc]} \BibitemShut
	{NoStop}%
	\bibitem [{\citenamefont {Martinez}\ and\ \citenamefont
		{Zanelli}(1996)}]{Martinez:1996gn}%
	\BibitemOpen
	\bibfield  {author} {\bibinfo {author} {\bibfnamefont {Cristian}\
			\bibnamefont {Martinez}}\ and\ \bibinfo {author} {\bibfnamefont {Jorge}\
			\bibnamefont {Zanelli}},\ }\bibfield  {title} {\enquote {\bibinfo {title}
			{{Conformally dressed black hole in (2+1)-dimensions}},}\ }\href {\doibase
		10.1103/PhysRevD.54.3830} {\bibfield  {journal} {\bibinfo  {journal} {Phys.
				Rev. D}\ }\textbf {\bibinfo {volume} {54}},\ \bibinfo {pages} {3830--3833}
		(\bibinfo {year} {1996})},\ \Eprint {http://arxiv.org/abs/gr-qc/9604021}
	{arXiv:gr-qc/9604021} \BibitemShut {NoStop}%
	\bibitem [{\citenamefont {Konoplya}(2004)}]{Konoplya:2004ik}%
	\BibitemOpen
	\bibfield  {author} {\bibinfo {author} {\bibfnamefont {R.A.}\ \bibnamefont
			{Konoplya}},\ }\bibfield  {title} {\enquote {\bibinfo {title} {{Influence of
					the back reaction of the Hawking radiation upon black hole quasinormal
					modes}},}\ }\href {\doibase 10.1103/PhysRevD.70.047503} {\bibfield  {journal}
		{\bibinfo  {journal} {Phys. Rev. D}\ }\textbf {\bibinfo {volume} {70}},\
		\bibinfo {pages} {047503} (\bibinfo {year} {2004})},\ \Eprint
	{http://arxiv.org/abs/hep-th/0406100} {arXiv:hep-th/0406100} \BibitemShut
	{NoStop}%
	\bibitem [{\citenamefont {Witten}(2020)}]{Witten:2020wvy}%
	\BibitemOpen
	\bibfield  {author} {\bibinfo {author} {\bibfnamefont {Edward}\ \bibnamefont
			{Witten}},\ }\bibfield  {title} {\enquote {\bibinfo {title} {{Matrix Models
					and Deformations of JT Gravity}},}\ }\href@noop {} {\  (\bibinfo {year}
		{2020})},\ \Eprint {http://arxiv.org/abs/2006.13414} {arXiv:2006.13414
		[hep-th]} \BibitemShut {NoStop}%
	\bibitem [{\citenamefont {Vieira}\ and\ \citenamefont
		{Bezerra}(2016)}]{Vieira:2016ubt}%
	\BibitemOpen
	\bibfield  {author} {\bibinfo {author} {\bibfnamefont {H.S.}\ \bibnamefont
			{Vieira}}\ and\ \bibinfo {author} {\bibfnamefont {V.B.}\ \bibnamefont
			{Bezerra}},\ }\bibfield  {title} {\enquote {\bibinfo {title} {{Confluent Heun
					functions and the physics of black holes: Resonant frequencies, Hawking
					radiation and scattering of scalar waves}},}\ }\href {\doibase
		10.1016/j.aop.2016.06.016} {\bibfield  {journal} {\bibinfo  {journal} {Annals
				Phys.}\ }\textbf {\bibinfo {volume} {373}},\ \bibinfo {pages} {28--42}
		(\bibinfo {year} {2016})},\ \Eprint {http://arxiv.org/abs/1603.02233}
	{arXiv:1603.02233 [gr-qc]} \BibitemShut {NoStop}%
	\bibitem [{\citenamefont {Fiziev}\ and\ \citenamefont
		{Staicova}(2011)}]{Fiziev:2011mm}%
	\BibitemOpen
	\bibfield  {author} {\bibinfo {author} {\bibfnamefont {Plamen}\ \bibnamefont
			{Fiziev}}\ and\ \bibinfo {author} {\bibfnamefont {Denitsa}\ \bibnamefont
			{Staicova}},\ }\bibfield  {title} {\enquote {\bibinfo {title} {{Application
					of the confluent Heun functions for finding the quasinormal modes of
					nonrotating black holes}},}\ }\href {\doibase 10.1103/PhysRevD.84.127502}
	{\bibfield  {journal} {\bibinfo  {journal} {Phys. Rev. D}\ }\textbf {\bibinfo
			{volume} {84}},\ \bibinfo {pages} {127502} (\bibinfo {year} {2011})},\
	\Eprint {http://arxiv.org/abs/1109.1532} {arXiv:1109.1532 [gr-qc]}
	\BibitemShut {NoStop}%
\end{thebibliography}

%

\end{document}